\documentclass[a4paper,11pt]{JHEP3}
\usepackage{amsmath}
\usepackage{epsf, amssymb}
\usepackage{epsfig}
\usepackage{slashed}

\newcommand{\Real}{{\rm Re}}
\newcommand{\Imagi}{{\rm Im}}
\newcommand{\half}{\frac{1}{2}}
\newcommand{\quart}{\frac{1}{4}}
\newcommand{\ie}{{\it i.e.\ }}
\newcommand{\eg}{{\it e.g.\ }}
\newcommand{\dT}{\int_{\cal F}\!\frac{d^2\tau}{(\Imagi\,\tau)^5}}
\newcommand{\dV}{\int_{\cal T}\prod_{r=1}^4 d^2v_r}
\newcommand{\dVdash}{\int_{\cal T}\prod_{r'=1}^4 d^2v_{r'}}
\newcommand{\chiprod}{\prod_{r<s}(\chi_{rs})^{\frac{1}{2}k_r \cdot k_s}}
\newcommand{\chiproddash}{\prod_{r'<s'}(\chi_{r's'})^{\frac{1}{2}k_{r'} \cdot k_{s'}}}
\newcommand{\ImTau}{\Imagi\,\tau}
\newcommand{\Imv}{\Imagi\,v}
\newcommand{\tr}{\mbox{Tr}}

\newcommand{\Rf}{{\mathcal{R}^4}}
\newcommand{\NSNS}{{$NS\otimes NS$}}
\newcommand{\RR}{{$R\otimes R$}}
\newcommand{\rem}[1]{}

\title{The One-Loop Five-Graviton Amplitude and the Effective Action}
\author{David M. Richards \\
        Department of Applied Mathematics and Theoretical Physics \\
        University of Cambridge \\
        Wilberforce Road, Cambridge CB3 0WA, United Kingdom.\\
        E-mail: \email{D.M.Richards@damtp.cam.ac.uk}}
\abstract{
We consider the one-loop five-graviton amplitude in type II string
theory calculated in the light-cone gauge. Although it is not possible
explicitly to evaluate the integrals over the positions of the vertex
operators, a low-energy expansion can be obtained, which
can then be used to infer terms in the low-energy effective
action. After subtracting diagrams due to known $D^{2n}R^4$ terms, we
show the absence of one-loop $R^5$ and $D^2R^5$ terms and determine
the exact structure of the one-loop $D^4R^5$ terms where,
interestingly, the coefficient in front of the $D^4R^5$ terms is
identical to the coefficient in front of the $D^6R^4$ term. Finally,
we show that, up to $D^6R^4\sim D^4R^5$, the $\epsilon_{10}$ terms
package together with the $t_8$ terms in the usual combination
$(t_8t_8\pm\frac{1}{8}\epsilon_{10}\epsilon_{10})$.
}
\preprint{DAMTP-2008-60}
\begin{document}

%%%%%%%%%%%%%%%%%%%%%%%%%%%%%%%%%%%%%%%%%%%%%%%%%%%%%%%%%%%%%%%%%%%%%%%%%%%%%%%%%%%%%%%%%%%%%%%%%%%%%%%%%%%%%%
\section{Introduction}
%%%%%%%%%%%%%%%%%%%%%%%%%%%%%%%%%%%%%%%%%%%%%%%%%%%%%%%%%%%%%%%%%%%%%%%%%%%%%%%%%%%%%%%%%%%%%%%%%%%%%%%%%%%%%%
At low energies, the way in which string theory differs from
conventional field theory is best encoded by the low-energy
effective action which, beyond lowest order, gives important stringy
corrections to supergravity. These corrections are relevant for a whole
host of physics. They modify Calabi-Yau compactifications to four
dimensions by, for example, correcting the metric for the universal
hypermultiplet \cite{Strominger:1997eb, Antoniadis:2003sw}. They are
also important for testing AdS/CFT beyond leading order, where they
give rise to $1/N$ and $1/\lambda$ effects in the field theory
\cite{Gubser:1998nz, Brodie:1998ke}. Further, if string theory is to
provide the microscopic description of black holes and black branes
then higher order corrections must play a crucial r\^ole
\cite{Mohaupt:2000mj, Kraus:2006wn, Castro:2008ne}. Stringy
corrections are also relevant for understanding the dualities
between string theories and eleven-dimensional supergravity
\cite{Russo:1997mk, Green:1997as}, and perhaps even for understanding
M-theory.

The meaning of the effective action in string theory is often not
well explained. It bears similarities to both the Wilsonian and 1PI
actions, but is identical to neither. The classical string field
theory action is a functional of both massless and massive fields,
$S(\phi_0,\phi_h)$. For low-energy physics only the massless modes are
explicitly relevant and so it makes sense to perform the path
integral over the massive modes,
\begin{equation}
  e^{iS_{\rm eff}(\phi_0)} = \int\mathcal{D}\phi_he^{iS(\phi_0,\phi_h)}.
\end{equation}
This differs from the usual Wilsonian effective action since the path
integral over massless modes with high-momentum has not been
performed. If amplitudes are calculated from $S_{\rm eff}(\phi_0)$ it
is still necessary to consider loops of massless particles. Generically
such an action will be non-local and its utility derives from a
low-energy expansion, which is equivalent to a derivative expansion.

Ideally we would be able to quantize $S(\phi_0,\phi_h)$ and find its
corresponding 1PI action, $\Gamma[\phi_{\rm cl}]$\footnote{This
assumes $\Gamma[\phi_{\rm cl}]$ exists. See \cite{Banks:2004xh} for
objections to this view.}. This would be some functional of all
possible fields in string theory, including massive fields and D-brane
fields.\footnote{Of course separating massless and massive fields is
ill-defined since certain massive fields become massless at special
points in moduli space.} Even in the absence of this, we can still
construct the one-particle irreducible action for $S_{\rm
eff}(\phi_0)$, which we call $S_{\rm eff,1PI}(\phi_0)$. This is
presumably equivalent to $\Gamma[\phi_{\rm cl}]$ with the massive
fields set to zero. The full string field theory 1PI action
$\Gamma[\phi_{\rm cl}]$ could in principle be used to find correlation
functions about any background of string theory. Since $S_{\rm
eff,1PI}(\phi_0)$ only involves fields which are massless in Minkowski
space, it is unclear whether it can be used in this way. However, it
may well still give amplitudes around backgrounds which share the same
massless spectrum as Minkowski space. For example, it is often
believed that $S_{\rm eff,1PI}(\phi_0)$ expanded in small fluctuations
around $AdS_5\times S^5$ gives correct $AdS_5\times S^5$ amplitudes
for massless states. However, it is not known whether correct results
would be obtained about backgrounds containing other massless fields,
such as various conifolds where D-branes become massless by wrapping
vanishing cycles.

Amplitudes derived from $S_{\rm eff,1PI}(\phi_0)$ should only include
tree diagrams. In this sense $S_{\rm eff,1PI}(\phi_0)$ resembles the
usual 1PI action. However, since massive fields are ignored and since
its status about backgrounds other than Minkowski is unclear, it is
perhaps not the full quantum effective action. The term effective
action in this paper, and in most of the string theory literature,
refers to $S_{\rm eff,1PI}(\phi_0)$, which we define as the action
whose tree diagrams generate string theory amplitudes in Minkowski
space at all orders in the string coupling\footnote{This is largely a
choice; there is no reason that we could not instead choose to
determine $S_{\rm eff}(\phi_0)$.}. However, it should be noted that
non-local terms due to thresholds will not be calculated. In
principle, this implies no loss of information since such terms can be
reinstated using unitarity and the tree-level effective action, but
without them the effective action cannot really be claimed to be the
full 1PI action.

There are at least three common ways to determine the effective
action. Firstly, one can calculate the world-sheet
$\beta$-functions. Since scale invariance at the quantum level
requires all $\beta$-functions to vanish, these lead to equations of
motion and an action for the background fields. Secondly, one can try
to use various symmetries, for example supersymmetry or
$SL(2,\mathbb{Z})$ invariance, to extend known terms to more complete
expressions. Finally, one can directly calculate string scattering
amplitudes and deduce the on-shell effective action which reproduces
them. It is this last method which is used here.

We calculate amplitudes using the light-cone gauge Green-Schwarz
formalism. The greatest problem with this formalism is that a
convenient representation for the vertex operators is only available
when all external states satisfy $k^+=0$. As a consequence, not all
terms in the amplitude, and so in the effective action, can be
discovered. For example, this formalism cannot be used to determine
whether the effective action contains the term
\begin{equation} \label{eq:ee1contraction}
  {\epsilon_{10\,m}}^{a_1\cdots a_9}{\epsilon_{10}}^{mb_1\cdots b_9}
  R_{a_1a_2b_1b_2} R_{a_3a_4b_3b_4} R_{a_5a_6b_5b_6}
  R_{a_7a_8b_7e} {{R_{a_9}}^e}{}_{b_8b_9},
\end{equation}
with only one contraction between the $\epsilon_{10}$
tensors. Similarly, the $B\wedge t_8R^4$ term found in type IIA
\cite{Vafa:1995fj} cannot be determined. The only terms that will be
missed in this paper are those with a single $\epsilon_{10}$ tensor
and those involving fewer than two contractions between a pair of
$\epsilon_{10}$ tensors. Terms like the famous
$\epsilon_{10}\epsilon_{10}R^4$, which has two contractions between
the $\epsilon_{10}$'s, will still be visible.

For both type II theories, the first correction to the
Einstein-Hilbert term is the tree-level $\alpha'^3R^4$
term\footnote{Powers of $\alpha'$ are relative to the Einstein-Hilbert
term, which itself contains an $\alpha'^{-4}$ factor.} which is given
in string frame by
\begin{equation} \label{eq:R4}
  \alpha'^3\int d^{10}x\sqrt{-g}\,e^{-2\phi}\,
  t_8^{a_1b_1\cdots a_4b_4}t_8^{c_1d_1\cdots c_4d_4}
  R_{a_1b_1c_1d_1}R_{a_2b_2c_2d_2}R_{a_3b_3c_3d_3}R_{a_4b_4c_4d_4},
\end{equation}
where $t_8$ is defined in the next section. There is a similar term at
the next order in string coupling with $e^{-2\phi}$ replaced by
$1$ and with a different coefficient. If terms involving
$\epsilon_{10}$ tensors are also considered then \eqref{eq:R4} is
extended by replacing $t_8t_8$ by
$(t_8t_8\pm\frac{1}{8}\epsilon_{10}\epsilon_{10})$. These terms are
supplemented by a whole host of terms involving fields other than the
graviton, such as $R^n(DH)^{4-n}$ \cite{Gross:1986mw}, $R^2(DF)^2$
\cite{Peeters:2003pv} and $\Lambda^{16}$ \cite{Green:1997me}, where
$H$ and $F$ are the \NSNS\ and \RR\ three-form field strengths
respectively and $\Lambda$ is the dilatino.

In the absence of an off-shell definition of string field theory, the
effective action is only fixed up to field redefinitions. As a
consequence, any terms which vanish when evaluated on the lower-order
equations of motion can be removed. In particular, this implies that
Ricci tensors and Ricci scalars can always be eliminated. So, at order
$\alpha'$ for example, a field redefinition can be used to remove
$R_{ab}R^{ab}$ and $R^2$, leaving just $R_{abcd}R^{abcd}$. The fact
that the Riemann-squared term also vanishes in type II is a
non-trivial consequence of this particular theory. Similarly, since
the Riemann and Weyl tensors only differ by terms involving $R_{ab}$
and $R$, the $t_8t_8R^4$ term can equivalently be rewritten as
$t_8t_8C^4$.

At higher orders in $\alpha'$ far less is known. Certain $D^{2n}R^4$
terms have been found from the expansion of the four-graviton
amplitude, for example $\alpha'^5t_8t_8D^4R^4$ and
$\alpha'^6t_8t_8D^6R^4$, but little is known about terms involving
more Riemann tensors or other fields. At $\alpha'^4$ and beyond it is
possible for $R^5$ and $D^{2n}R^5$ terms to appear. The purpose of
this paper is to determine such terms at one-loop up to and including
$\alpha'^6D^4R^5$. To do so we calculate the five-graviton amplitude
on a toroidal world-sheet and determine its expansion in powers of $\alpha'$. Before
new terms can be constructed, it is important to subtract
contributions from known $D^{2n}R^4$ terms. After doing so, the
remaining terms (if any) can be covariantised to give new $R^5$ and
$D^{2n}R^5$ terms. The plan of this paper is as follows. Section
\ref{sec:4grav} reviews the case of four-gravitons -- the amplitude,
its expansion and the effective action -- which leads to the
well-known $D^{2n}R^4$ terms. The calculation of the five-graviton
amplitude is reviewed and simplified in section \ref{sec:5grav} before
determining its low-energy expansion in section \ref{sec:5gravExp}
using an extension of the four-graviton techniques. Section
\ref{sec:5gravEff} calculates all the relevant field theory diagrams
arising from the known $D^{2n}R^4$ terms. After subtracting these from
the expanded amplitude, the presence of $R^5$ and $D^{2n}R^5$ terms
are determined. The extension of this analysis to include
$\epsilon_{10}\epsilon_{10}$ terms is contained in section
\ref{sec:5gravEps}, where the equivalent of
$(t_8t_8\pm\frac{1}{8}\epsilon_{10}\epsilon_{10})R^4$ at higher orders
in $\alpha'$ is studied. Appendix \ref{sec:tensors} contains various
identities linking the three tensors, $t_{10}$, $\bar{t}_{10}$ and
$t_8$, which arise when calculating the amplitude. Throughout we will
use a metric with signature $\{-,+,+,+,\ldots\}$ and will often set
$2\alpha'=1$.

%%%%%%%%%%%%%%%%%%%%%%%%%%%%%%%%%%%%%%%%%%%%%%%%%%%%%%%%%%%%%%%%%%%%%%%%%%%%%%%%%%%%%%%%%%%%%%%%%%%%%%%%%%%%%%
\section{The Effective Action from the Four-Graviton Amplitude} \label{sec:4grav}
\begin{figure}
\begin{center}
\includegraphics[scale=0.5]{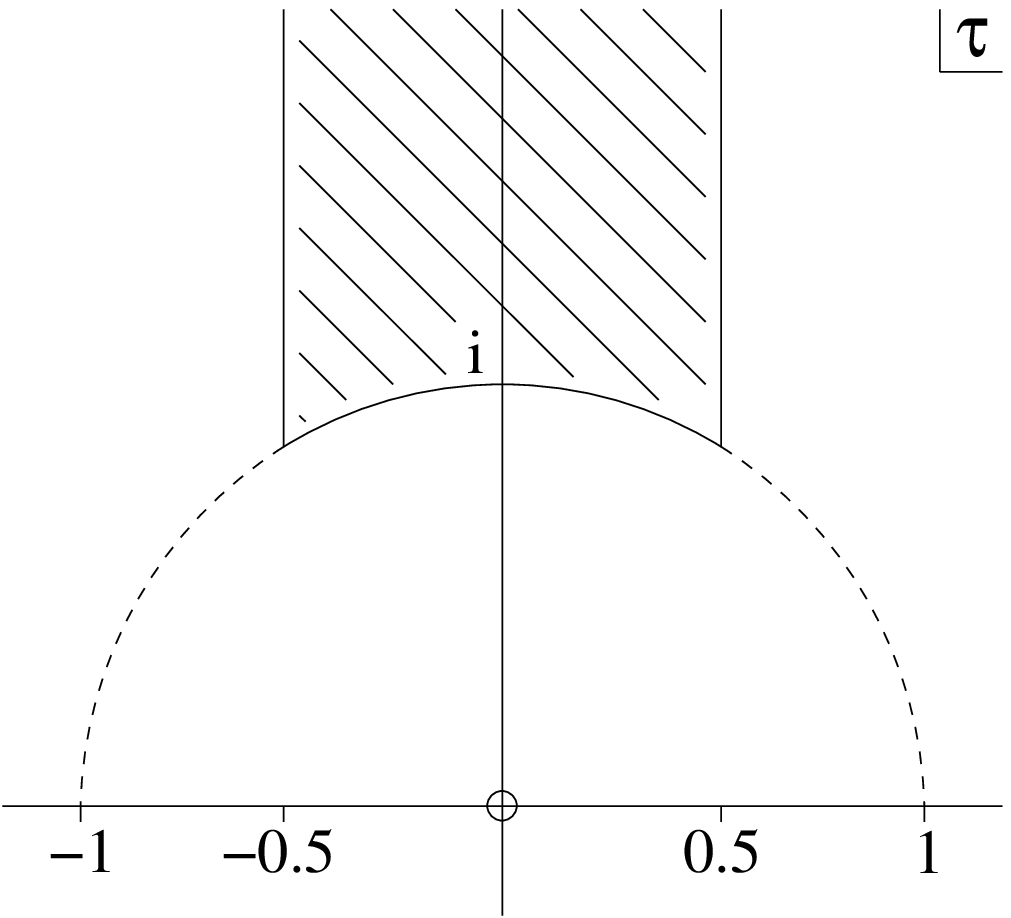}
\caption{One of the fundamental domains of $SL(2,\mathbb{Z})$.}
\label{fig:funddom}
\end{center}
\end{figure}

%%%%%%%%%%%%%%%%%%%%%%%%%%%%%%%%%%%%%%%%%%%%%%%%%%%%%%%%%%%%%%%%%%%%%%%%%%%%%%%%%%%%%%%%%%%%%%%%%%%%%%%%%%%%%%
Before we embark on studying the type II five-graviton amplitude, we
review the equivalent results for the four-graviton amplitude. Let
the four particles have polarisation tensors $h^r_{a_rb_r}$ and
momenta $k^r_{a_r}$, where $r=1, \ldots, 4$ and $a, b=0, \ldots,
9$. There are three Mandelstam variables defined by
\begin{equation}
  s=-(k_1+k_2)^2,\qquad t=-(k_1+k_3)^2,\qquad u=-(k_1+k_4)^2,
\end{equation}
which are related by $s+t+u=0$. The one-loop amplitude is well-known
to be given by \cite{Green:1981xx}
\begin{equation} \label{eq:4gravAmp}
   A_{4h} = \hat{K} \dT \int_{\cal T}\prod_{r=1}^3 d^2v_r \chiprod,
\end{equation}
where
\begin{equation}
  \hat{K} = t_8^{a_1b_1\cdots a_4b_4} t_8^{c_1d_1\cdots c_4d_4}
  k^1_{a_1}k^1_{c_1}h^1_{b_1d_1} k^2_{a_2}k^2_{c_2}h^2_{b_2d_2}
  k^3_{a_3}k^3_{c_3}h^3_{b_3d_3} k^4_{a_4}k^4_{c_4}h^4_{b_4d_4},
\end{equation}
and where we have set $2\alpha'=1$.
Here $v_r$ are the positions of the four vertex operators on the
world-sheet torus and are integrated over the rectangular region
$-\half<\Real\,v\leq\half, -\half\ImTau<\Imv\leq\half\ImTau$, which
we denote by ${\cal T}$. The
variable $\tau$ parameterizes the modulus of the torus and so should
be integrated over a fundamental domain of $SL(2,\mathbb{Z})$; to
facilitate the low-energy expansion, it is convenient to use the
fundamental domain given in figure \ref{fig:funddom}, which we denote
by ${\cal F}$. The function
$\chi_{rs}\equiv\chi(v_{rs},\tau)$, $v_{rs}\equiv v_r-v_s$, is a
non-singular doubly periodic function of $v$ and $\bar{v}$ which is
explicitly given by
\begin{equation}
  \chi(v,\tau) = 2\pi \exp \left(
  -\frac{\pi(\Imagi\,v)^2}{\ImTau} \right)
  \left|\frac{\theta_1(v,\tau)}{\theta_1'(0,\tau)}\right|,
\end{equation}
where $\theta_1(v,\tau)$ is the usual Jacobi theta function given, for
example, in \cite{Green:1987mn}.

The $t_8$ tensor is an eight-component tensor originating from the
trace over eight fermionic zero modes and can be written explicitly as
a sum of an eight-component $\epsilon_8$ tensor and sixty
$\delta\delta\delta\delta$ tensors \cite{Green:1987mn},
\begin{align} \label{eq:t8tensor}
  t_8^{a_1b_1a_2b_2a_3b_3a_4b_4} =&
  \pm{\textstyle\half}\epsilon_8^{a_1b_1a_2b_2a_3b_3a_4b_4} \notag\\
  & -{\textstyle\half}\big(
  (\delta^{a_1a_2}\delta^{b_1b_2}-\delta^{a_1b_2}\delta^{a_2b_1})
  (\delta^{a_3a_4}\delta^{b_3b_4}-\delta^{a_3b_4}\delta^{a_4b_3}) \notag\\
  & \qquad + (1,2,3,4) \to (1,3,2,4) \notag\\
  & \qquad + (1,2,3,4) \to (1,4,2,3) \big) \notag\\
  & +{\textstyle\half}\big( \delta^{a_1a_2}\delta^{b_1b_3}\delta^{b_2b_4}\delta^{a_3a_4}
    + \delta^{a_1a_2}\delta^{b_1b_4}\delta^{b_2b_3}\delta^{a_3a_4}
  + \delta^{a_1a_3}\delta^{b_1b_4}\delta^{b_2b_3}\delta^{a_2a_4} \notag\\
  & \qquad + \mbox{45 other terms determined by antisymmetry} \big),
\end{align}
with the $\pm$ sign depending on the $SO(8)$ chirality. It has
symmetries not dissimilar to the Riemann tensor: it is antisymmetric
under interchange of any pair of indices $a_r$ with $b_r$, and is
symmetric under interchanging the pair $(a_r,b_r)$ with the pair
$(a_s,b_s)$. It is worth noting that $t_8$ is often defined without
the $\epsilon_8$ tensor, especially in effective actions, and we will
clarify this issue later. However, for the four-graviton amplitude
this difference is unimportant since all terms involving an $\epsilon$
vanish due to momentum conservation.

As with all amplitudes, $A_{4h}$ is not necessarily finite for all
values of the external momenta. Poles occur when the momenta are
such that an on-shell intermediate particle can be produced. On the
world-sheet, this is interpreted as two states approaching each other
and developing a long tube separating them from the other
states. By expanding around $v_{rs}=0$ for some fixed $r$, $s$, and
using $\chi(v,\tau) \sim 2\pi|v|$ for small $v$, it is easy to show
that the lightest pole goes like $\frac{1}{\alpha'k_r\cdot k_s+2}$ and
so is massive. The absence of massless poles is consistent with the
vanishing of the one-loop amplitude for three gravitons. There are
also threshold branch cuts when the external momenta are sufficiently
large to produce two or more physical states which circulate around
the loop. These originate from the region where the torus degenerates
to a thin wire, \ie\ $\ImTau\to\infty$, which causes $\chi(v,\tau)$ to
diverge.

However, it is not possible to see all these poles and thresholds from
the integral representation of the amplitude given in
\eqref{eq:4gravAmp}. In fact, the integral representation only
converges at the single point $s=t=u=0$. The resolution is to split
the integration over the $v_r$ into six regions depending on the
ordering of the $\Imagi\,v_r$. For example, for the region $0 \leq
\Imagi\,v_1 \leq \Imagi\,v_2 \leq \Imagi\,v_3 \leq \Imagi\,\tau$, we
eliminate $t$ and consider the amplitude as a function of complex $s$
and $u$. Then it can be shown that avoiding singularities when
$v_{rs}\to 0$ requires $s<8$, $u<8$ and $s+u>-8$. Similarly, avoiding
the singularity as $\ImTau\to\infty$ requires $s\leq 0$ and $u\leq
0$. So, taken together, this region of the amplitude only converges in
the infinite strip $s\leq 0$, $u\leq 0$ and $s+u>-8$. It is then
possible to analytically continue this strip to the entire complex
plane. The real physical amplitude should be understood as the sum of
the continuation of all six regions. Only after continuing can the
amplitude be shown to contain all the correct massive poles, massive
double poles and thresholds required by unitarity
\cite{D'Hoker:1993ge, D'Hoker:1993mr, D'Hoker:1994yr}.

\subsection{Low-energy Expansion of the Amplitude}
Before determining the low-energy effective action, it is necessary to
expand the amplitude in powers of $\alpha's$, $\alpha't$ and
$\alpha'u$. In particular, we need to expand
\begin{equation} \label{eq:4gravI}
  I = \dT \int_{\cal T}\prod_{r=1}^3 d^2v_r \chiprod.
\end{equation}
Massive poles in $A_{4h}$ will not be visible at low energies, but
threshold corrections will still be present as branch cuts. Due to
the symmetry of $I$ in $s$, $t$ and $u$, the expansion will have the
form,
\begin{equation} \label{eq:4gravExp}
  I = a + \frac{\alpha'}{4}I_{\rm nonan}(s,t,u) +
  b\frac{\alpha'^2}{16}(s^2+t^2+u^2) +
  c\frac{\alpha'^3}{64}(s^3+t^3+u^3) + \cdots,
\end{equation}
where $a$, $b$ and $c$ are constants that we wish to
determine. $I_{\rm nonan}$ is the first non-analytic term due to
thresholds and contains terms of the form $s\ln s$. Using unitarity, it
can be seen to arise from two four-graviton tree-level amplitudes
connected by a pair of on-shell gravitons. The lack of other
non-analytic terms before order $\alpha'^4$ is related to the fact
that, for type II theories, the first correction to
the effective action is at order
$\alpha'^3$ relative to the Einstein-Hilbert term.
There is no analytic term at order $\alpha'$
since $s+t+u$ vanishes. Despite being symmetric in $s$, $t$, $u$, there is no need for
an $st+su+tu$ term at order $\alpha'^2$ since this is proportional to
$s^2+t^2+u^2$. Similarly, at order $\alpha'^3$, the other possible symmetric
expressions, \ie\ $st^2+su^2+\ldots$ and $stu$, are both proportional
to $s^3+t^3+u^3$. It is worth noting that this property does not
continue indefinitely: eventually terms other than $s^n+t^n+u^n$ will
appear. Since $s$, $t$ and $u$ are not independent it is useful to
eliminate $u$ leaving,
\begin{equation} \label{eq:4gravExp2}
  I-I_{\rm nonan} = a + b\frac{\alpha'^2}{8}(s^2+t^2+st) -
  c\frac{3\alpha'^3}{64}(s^2t+st^2) + \cdots.
\end{equation}

The constants $a$, $b$ and $c$ were first determined in
\cite{Green:1999pv}, which we review here since similar techniques
will be required to expand the five-graviton amplitude. By
differentiating \eqref{eq:4gravExp2} and taking the limit $s,t\to 0$,
we need to consider expressions such as
\begin{equation} \label{eq:cConstant}
  c = -\frac{2^8}{3(2\alpha')^3} \lim_{s,t\to 0}
  \partial_s^2\partial_t (I-I_{\rm nonan}),
\end{equation}
where
\begin{align} \label{eq:ModIntDerivatives}
  \lim_{s,t\to 0}\partial_s^m\partial_t^n I &= \left(-\quart\right)^{m+n}\dT
  \int_{\cal T}\prod_{r=1}^3
  d^2v_r\,(\ln\chi_{12}+\ln\chi_{34}-\ln\chi_{14}-\ln\chi_{23})^m
  \notag \\
  &\qquad\quad\times (\ln\chi_{13}+\ln\chi_{24}-\ln\chi_{14}-\ln\chi_{23})^n.
\end{align}
The non-analytic parts originate from the region of moduli space
where $\ImTau\to\infty$, interpreted as the torus degenerating into a thin
wire. To remove them, the $\tau$-integral over the fundamental
domain is split into two parts: a part with $\ImTau\le L$, for large
$L$, which will contain  the required
constant term and an $L$-dependent term; and a part with $\ImTau>L$,
which will contain the non-analytic piece and an $L$-dependent
term. Since the overall integral cannot depend on $L$, the
$L$-dependent terms must cancel between the two regions. The constant
piece can be found by restricting to the first region and ignoring
$L$-dependent pieces. To perform the $v_r$-integrals, it is convenient
to represent $\ln\chi$ as a Fourier series,
\begin{align} \label{eq:lnchi}
  \ln\chi(v,\tau) &=
  -\frac{1}{2\pi}\sum_{(m,n)\neq(0,0)}\frac{\ImTau}{|m\tau+n|^2} \exp\left[ 2\pi im\left(\Real\,v -
  \frac{\Real\,\tau}{\ImTau}\Imagi\,v\right) -2\pi
  in\,\frac{\Imagi\,v}{\ImTau} \right] \notag \\
  &\qquad\quad- \left|\ln\sqrt{2\pi}\eta(\tau) \right|^2,
\end{align}
where $\eta(\tau)$ is the Dedekind eta-function. Since $\ln\chi$ always occurs
in positive and negative pairs in \eqref{eq:ModIntDerivatives}, the
zero mode part containing $\eta(\tau)$ cancels out and can be
ignored.

Since there are no non-analytic terms at lowest order, the calculation
of $a$ can be performed by integrating $\tau$ over the entire
fundamental domain,
\begin{equation}
  a = \lim_{s,t\to 0} I = \dT \int_{\cal T}\prod_{r=1}^3 d^2v_r =
  \int_{\cal F}\!\frac{d^2\tau}{(\ImTau)^2} = \frac{\pi}{3}.
\end{equation}

\subsubsection{Calculation of $b$} \label{sec:4gravB}
Setting $2\alpha'=1$ and twice differentiating \eqref{eq:4gravExp2}
with respect to $s$, the coefficient $b$ is given by
\begin{equation}
  b = \dT \int_{\cal T}\prod_{r=1}^3 d^2v_r\,
    (\ln\chi_{12}+\ln\chi_{34}-\ln\chi_{14}-\ln\chi_{23})^2\,\bigg|_{\rm const},
\end{equation}
where $|_{\rm const}$ refers to the constant ($L$-independent) piece
of the restricted $\tau$-integral, \ie\ the integral over only the
$\ImTau\le L$ region. Expanding the brackets yields two type of terms:
cross-terms and square-terms. After redefinitions of the $v_r$, the
$v_r$-integrals for all six of the cross-terms are equivalent to
$\left(\int_{\cal T} d^2v\,\ln\chi\right)^2 \left(\int_{\cal T} d^2v\right)$, which
vanishes since the integral over a single $\ln\chi(v,\tau)$ is zero
(recall that the zero mode part of $\ln\chi$ has been removed). The
four square-terms are all equivalent to
\begin{equation}
  \left(\int_{\cal T} d^2v\,(\ln\chi)^2\right) \left(\int_{\cal T} d^2v\right)^2 =
  (\ImTau)^2\int_{\cal T} d^2v\,(\ln\chi)^2.
\end{equation}
The remaining integral over $v$ is easily performed using
\eqref{eq:lnchi},
\begin{equation}
  \int_{\cal T} d^2v\,(\ln\chi)^2 =
  \frac{\ImTau}{4\pi^2}\sum_{(m,n)\neq(0,0)}\frac{(\ImTau)^2}{|m\tau+n|^4},
\end{equation}
which leaves $b$ as an integral over $\tau$ where, as discussed above,
the non-analytic part can be removed by restricting the integral to
the lower region of the fundamental domain,
\begin{equation}
  b_L = \frac{1}{\pi^2}\int_{\ImTau<L}\!\frac{d^2\tau}{(\ImTau)^2}
  \sum_{(m,n)\neq(0,0)}\frac{(\ImTau)^2}{|m\tau+n|^4},
\end{equation}
and ignoring the $L$-dependent pieces. Generically, modular integrals
are difficult to evaluate. However, in this case, the sum can be
identified as the Epstein zeta function $Z_2(\tau,\bar\tau)$, obeying
the equation $Z_2=2(\ImTau)^2\partial_{\tau}\partial_{\bar{\tau}}Z_2$,
which makes the integrand a total derivative and reduces the
calculation of $b_L$ to an integral over the boundary. Due to the
identification under $SL(2,\mathbb{Z})$, the fundamental domain in
figure \ref{fig:funddom} should be thought of as rolled-up into a
cigar, with the only boundary at $\ImTau=L$. Using that
$Z_2(\tau,\bar\tau)\sim\pi^4(\ImTau)^2/45$ for large $\ImTau$, it is
found that
\begin{equation}
  b_L = \frac{2^4\pi^2}{6!}L + \mathcal{O}(L^{-2}).
\end{equation}
The $L$-dependence must, as confirmed in \cite{Green:1999pv},
cancel with the $L$-dependent piece from the region with
$\ImTau>L$. Since there is no $L$-independent piece, we conclude that
$b=0$.

\subsubsection{Calculation of $c$} \label{sec:4gravC}
From \eqref{eq:cConstant}, the calculation of $c$ reduces to
\begin{align}
  c &= \frac{4}{3} \dT \int_{\cal T}\prod_{r=1}^3 d^2v_r\,
  (\ln\chi_{12}+\ln\chi_{34}-\ln\chi_{14}-\ln\chi_{23})^2 \notag\\
    &\qquad\quad\times (\ln\chi_{13}+\ln\chi_{24}-\ln\chi_{14}-\ln\chi_{23})
    \,\Big|_{\rm const}.
\end{align}
After expanding the brackets, there are two types of potentially
non-zero integrals: integrals involving
$\ln\chi_{ab}\ln\chi_{bc}\ln\chi_{ca}$ and integrals involving
$(\ln\chi_{ab})^3$; all other types vanish due to the vanishing of
single $\ln\chi(v,\tau)$ integrals. The $v$-integrals in the first
kind give
\begin{equation}
  -\frac{(\ImTau)^2}{8\pi^3}\sum_{(m,n)\neq(0,0)}\frac{(\ImTau)^3}{|m\tau+n|^6}
  \equiv -\frac{(\ImTau)^2}{8\pi^3}Z_3(\tau,\bar\tau),
\end{equation}
where $Z_3(\tau,\bar\tau)$ is another non-holomorphic Eisenstein series, satisfying a
similar equation to $Z_2(\tau,\bar\tau)$. Again the
integral over $\tau$ reduces to an integral over the boundary of the
fundamental domain and, as with
the evaluation of $b$, this only contains an $L$-dependent piece,
which makes no contribution to the constant
$c$. The other type of $v$-integral gives
\begin{equation} \label{eq:lnchi3}
  \int_{\cal T} d^2v\,(\ln\chi)^3 =
  -\frac{\ImTau}{8\pi^3}\sum_{(m,n),(k,l),(p,q)\neq(0,0)}
  \frac{(\ImTau)^3\,\delta_{m+k+p}\delta_{n+l+q}}{|m\tau+n|^2|k\tau+l|^2|p\tau+q|^2},
\end{equation}
which must now be integrated over $\tau$. However, the right-hand-side
is now not an Epstein zeta function and so the integral cannot
obviously be evaluated by writing the integrand as a total
derivative. Instead, \cite{Green:1999pv} makes use of an `unfolding
procedure', which is not reviewed here, where $\ln\chi$ is represented
as a Poincar\'e series which converts the integral over the complex
plane to an integral over the semi-infinite line. Again, it is found
that the $L$-dependent piece cancels with the $L$-dependent piece from
the same integral over the upper fundamental domain. However, there is
now also an $L$-independent contribution, which leads to a non-zero
value for $c$,
\begin{equation}
  c = \frac{2}{3\pi}\zeta(2)\zeta(3).
\end{equation}
Collecting the results for $a$, $b$ and $c$, we can write the
low-energy expansion of $A_{4h}$ up to order $\alpha'^3$ as
\begin{equation} \label{eq:4gravExpFinal}
  \hat{K}(I - I_{\rm nonan}) = \hat{K}\left(\frac{\pi}{3} +
  \frac{\zeta(3)\pi\alpha'^3}{3^2\cdot 2^6}(s^3+t^3+u^3) +
  \mathcal{O}(\alpha'^4)\right).
\end{equation}

\subsection{The Effective Action} \label{sec:4gravEffAct}
Terms in the type II effective action can be deduced from the
covariantisation of the expansion \eqref{eq:4gravExpFinal}. The first
such term occurs at order $\alpha'^3$ and is the one-loop partner of
the tree-level result found in \cite{Gross:1986iv}. Combining the
tree-level and one-loop results, it is given in string frame by
\begin{equation}
  \alpha'^3\int d^{10}x \sqrt{-g}\,e^{-\phi/2}\left(2\zeta(3)e^{-3\phi/2} +
  2\pi\cdot\frac{\pi}{3}e^{\phi/2}\right) \Rf,
\end{equation}
where $\Rf$ is shorthand for
\begin{equation}
  t_8^{a_1b_1a_2b_2a_3b_3a_4b_4}t_8^{c_1d_1c_2d_2c_3d_3c_4d_4}
  R_{a_1b_1c_1d_1}R_{a_2b_2c_2d_2}R_{a_3b_3c_3d_3}R_{a_4b_4c_4d_4},
\end{equation}
and where the normalisation has been chosen so that the coefficients
agree with those in \cite{Green:2005ba}. At the next order, the
expansion vanishes by momentum conservation. However, this cannot be
used to conclude that there is no one-loop $t_8t_8D^2R^4$ term since
such a term would not contribute to the four-graviton amplitude. In
the next chapter we will use the five-graviton amplitude to
demonstrate that such a term really is absent. At order $\alpha'^5$
the expansion again vanishes, but this time it does imply the absence
of the one-loop $D^4R^4$ term, so that combining with the non-zero
tree-level contribution gives
\begin{equation}
  \alpha'^5\int d^{10}x \sqrt{-g}\,e^{\phi/2}\left(2\zeta(5)e^{-5\phi/2} +
  0\cdot e^{-\phi/2}\right) D^4\Rf.
\end{equation}
Finally, at order $\alpha'^6$, the non-zero value for $c$ leads to a
new $D^6R^4$ term,
\begin{equation}
  \alpha'^6\int d^{10}x \sqrt{-g}\,e^{\phi}\left(4\zeta(3)^2e^{-3\phi} +
  8\zeta(2)\zeta(3)e^{-\phi}\right) D^6\Rf.
\end{equation}
It is not possible using the four-graviton amplitude to determine
exactly how the derivatives act on the Riemann tensors. For example,
at order $D^4R^4$, the difference between $(D_eD_fR)(D^eD^fR)R^2$ and
$(D_eD^eD_fD^fR^2)R^2$ cannot be distinguished. The five-graviton
amplitude will allow some of these issues to be resolved.

\subsubsection{Complete Coupling Dependence in IIB} \label{sec:IIBEffAct}
This summarises the known tree-level and one-loop results up to order
$\alpha'^6$, but it is interesting to ask about results at higher
orders in the string coupling. Because of the complexity involved,
there has been little progress in directly calculating higher-loop
amplitudes, although some results have been found at two-loops
(\cite{D'Hoker:2002gw, D'Hoker:2005jb, D'Hoker:2005jc, D'Hoker:2007ui,
Berkovits:2004px}). However, using various other techniques, such as
compactifications of 11D supergravity, supersymmetry and
$SL(2,\mathbb{Z})$ invariance, it is possible in IIB to find certain
all-order expressions, even including non-perturbative effects. For
example, it was shown in \cite{Green:1997tv, Green:1997di,
Green:1998by, Green:1997as} that the complete $\Rf$ action is given by
\begin{equation}
  \alpha'^3\int d^{10}x \sqrt{-g}\,e^{-\phi/2}Z_{3/2}(\tau,\bar\tau) \Rf,
\end{equation}
where $Z_{3/2}$ is a non-holomorphic Eisenstein series given by
\begin{align}
  Z_{3/2}(\tau,\bar\tau) &= \sum_{(m,n)\neq(0,0)}
  \frac{(\ImTau)^{3/2}}{|m\tau+n|^3} \notag\\
  &= 2\zeta(3)e^{-3\phi/2} +
  \frac{2\pi^2}{3}e^{\phi/2} \notag\\
  &\qquad + 4\pi\sum_{k\neq
  0}\mu(k)e^{-2\pi(|k|e^{-\phi}-ikC^{(0)})}k^{1/2}
  (1+\frac{3}{16\pi|k|}e^{\phi}+\cdots),
\end{align}
with $\mu(k)=\sum_{d|k}d^{-2}$.
Here $\tau$, which should not be confused with the modular parameter
in one-loop amplitudes, is the usual combination of the Ramond-Ramond
scalar and the dilaton, $\tau = C^{(0)} + ie^{-\phi}$. The expansion
shows that there are no perturbative contributions beyond tree-level
and one-loop, but that there are an infinite sum of single D-instanton
terms, which were first studied in \cite{Green:1997tv}.

Using the two-loop 11D supergravity four-graviton amplitude, it
was suggested in \cite{Green:1999pu} that the equivalent result at
order $\alpha'^5$ should be
\begin{equation}
  \alpha'^5\int d^{10}x \sqrt{-g}\,e^{\phi/2}Z_{5/2}(\tau,\bar\tau) D^4\Rf,
\end{equation}
with $Z_{5/2}(\tau,\bar\tau)$ another non-holomorphic Eisenstein
series given by
\begin{align} \label{eq:IIBD4R4}
  Z_{5/2}(\tau,\bar\tau) &= \sum_{(m,n)\neq(0,0)}
  \frac{(\ImTau)^{5/2}}{|m\tau+n|^5} \notag\\
  &= 2\zeta(5)e^{-5\phi/2} + 0\cdot e^{-\phi/2} +
  \frac{4\pi^4}{45}e^{3\phi/2} \notag\\
  &\qquad + \frac{8\pi^2}{3}\sum_{k\neq
  0}\mu'(k)e^{-2\pi(|k|e^{-\phi}-ikC^{(0)})}k^{3/2}
  \left(1+\frac{15}{16\pi|k|}e^{\phi}+\cdots\right),
\end{align}
where $\mu'(k)=\sum_{d|k}d^{-4}$. In addition to the tree and one-loop
terms, there is now a prediction for a non-zero perturbative term at
two-loops, which was confirmed by direct calculation in
\cite{D'Hoker:2005ht, Zheng:2002um, Zheng:2002uu, Zheng:2002ji}. There
are no further perturbative corrections but, as with the $\Rf$ case,
there is an infinite sum due to single D-instantons.

Finally, the $D^6\Rf$ case was first studied in \cite{Green:2005ba},
where it was conjectured to be given by
\begin{equation}
  \alpha'^6\int d^{10}x\sqrt{-g}\,e^{\phi}
  \mathcal{E}_{\frac{3}{2},\frac{3}{2}}(\tau,\bar\tau) D^6\Rf,
\end{equation}
where
\begin{align} \label{eq:IIBD6R4}
  \mathcal{E}_{\frac{3}{2},\frac{3}{2}}(\tau,\bar\tau) &=
  4\zeta(3)^2e^{-3\phi} + 8\zeta(2)\zeta(3)e^{-\phi} +
  \frac{48\zeta(2)^2}{5}e^{\phi} +
  \frac{32\zeta(2)\zeta(4)}{63}e^{3\phi} \notag\\
  & \qquad+ \sum ( \mbox{single D-insts} + \mbox{double D-insts} ).
\end{align}
The tree and one-loop coefficients again agree with the expansions of
the equivalent amplitudes, and there are two- and three-loop
predictions, neither of which has been confirmed by direct
calculation. Non-perturbatively there are now infinite sums of both
single D-instantons and pairs of D-instantons.

%%%%%%%%%%%%%%%%%%%%%%%%%%%%%%%%%%%%%%%%%%%%%%%%%%%%%%%%%%%%%%%%%%%%%%%%%%%%%%%%%%%%%%%%%%%%%%%%%%%%%%%%%%%%%%
\section{The Five-Graviton Amplitude} \label{sec:5grav}
%%%%%%%%%%%%%%%%%%%%%%%%%%%%%%%%%%%%%%%%%%%%%%%%%%%%%%%%%%%%%%%%%%%%%%%%%%%%%%%%%%%%%%%%%%%%%%%%%%%%%%%%%%%%%%
The light-cone gauge, GS formalism, five-graviton one-loop amplitude
was first calculated in \cite{Frampton:1986gi, Lam:1986kg}. However,
the intention was only to demonstrate modular invariance and there was
no attempt to simplify the amplitude. Here we review the calculation
and present simplifications necessary for extracting the effective
action in subsequent chapters.

\subsection{Calculating the Amplitude} \label{sec:amp}
Let the five gravitons have
polarisation tensors $h^r_{a_rb_r}$ and momenta $k^r_{a_r}$, where now
$r=1, \ldots, 5$; since we are in light-cone gauge $a$ and $b$ range
from $1$ to $8$. The graviton vertex operator is given by
\cite{Green:1987sp}
\begin{equation}
  \mathcal{V}_h(k,z) = h_{ac} (\partial X^a(z) - R^{ab}(z)k^b) (\bar\partial X^c(z) -
  \tilde{R}^{cd}(z)k^d) e^{ik\cdot X(z)},
\end{equation}
where $R^{ab}(z)\equiv\quart S^A(z)\gamma_{AB}^{ab}S^B(z)$. $X^a(z)$,
$S^A(z)$ and $\tilde{S}^A(z)$ are the bosonic and fermionic string
coordinates. Motivated by the usual prescription for calculating GS
amplitudes, explained in \cite{Green:1987mn}, we consider
\begin{equation} \label{eq:5gravAmp}
  A_{5h} = \int_{\cal F}\!d^2\tau \dV \int d^{10}p\,\tr
  \left(\mathcal{V}_h(k_1,\rho_1)\cdots\mathcal{V}_h(k_5,\rho_5)
  w^{L_0}\bar{w}^{\tilde L_0}\right),
\end{equation}
where $v_r = \ln\rho_r/2\pi i$, $\tau = \ln w/2\pi i$, $w=\rho_5$, and
the trace is over all $\alpha_n$, $\tilde\alpha_n$, $S_n$ and
$\tilde{S}_n$ modes. Here $\alpha_n$ and $\tilde{\alpha}_n$ are the
left- and right-moving bosonic modes, and $S_n$ and $\tilde{S}_n$ are
the left- and right-moving fermionic modes. $L_0 = \frac{1}{8}p^2 +
\sum_{n>0}(\alpha_{-n}^a\alpha_n^a + nS_{-n}^AS_n^A)$ is the
left-moving zeroth Virasoro generator; the right-moving equivalent,
$\tilde L_0$, has a similar expression. The trace over $S$ vanishes
unless there are at least eight $S_0$ zero modes (and similarly for
$\tilde{S}$) and so there are only three types of term to consider:
one term containing $R^5\tilde{R}^5$, ten terms containing
$\partial{X}R^4\tilde{R}^5$ or $R^5\bar\partial{X}\tilde{R}^4$, and
twenty-five terms containing
$\partial{X}R^4\bar\partial{X}\tilde{R}^4$.

\subsubsection{The Term Containing $R^5\tilde{R}^5$}
Since this term contains no $\partial X$ or $\bar\partial X$ factors,
the trace over the $\alpha$ and $\tilde\alpha$ modes and the $p$-integral
give the same $|f(w)|^{-16} \left(\frac{2}{\ImTau}\right)^5 \chiprod$
factor found in the four-graviton amplitude, where $f(w) \equiv
\prod_{n=1}^{\infty} (1-w^n)$. The trace over the $S$ modes involves
five products of
\begin{equation}
  R^{ab}(z) = \quart\Big(S_0^A+\sum_{m\neq 0}S_m^Az^m\Big) \gamma_{AB}^{ab}
  \Big(S_0^B+\sum_{n\neq 0}S_n^Bz^n\Big)
\end{equation}
and, since the trace over nine $S_0$ modes vanishes, there are only two
types of contributions: a term with ten $S_0$ modes, and a term with
eight $S_0$ modes and two non-zero $S$ modes. The trace over ten $S_0$
modes leads to a ten-index tensor,
\begin{equation} \label{eq:t10}
  t_{10}^{a_1b_1\cdots a_5b_5} = tr_{S_0}(R_0^{a_1b_1}\cdots R_0^{a_5b_5}),
\end{equation}
which can be written as a sum of forty $t_8\delta$ tensors, as given
in \eqref{eq:t10tot8}. In the second case, the non-zero modes cannot
come from the same $R^{ab}$ tensor since their trace will lead to a
vanishing $\delta^{AB}\gamma_{AB}$ factor, so we only need consider
terms like
\begin{equation} \label{eq:R5R5trace}
  4\,\tr_{S_0}(R_0^{a_1b_1}R_0^{a_2b_2}R_0^{a_3b_3} {\textstyle\quart}
  S_0^A\gamma_{AB}^{a_4b_4} {\textstyle\quart} S_0^C\gamma_{CD}^{a_5b_5})
  \times\tr_{S_{\slashed{0}}}\Big(\sum_{n\neq 0} S^B_n z^n
  \sum_{m\neq 0} S^D_m z'^{\,m} w^{\sum pS_{-p}\cdot S_p}\Big),
\end{equation}
where $S_{\slashed{0}}$ represents the non-zero modes of $S$, \ie $S_n$
with $n\neq 0$. The trace over these non-zero $S$ modes gives
$\delta^{BD}f(w)^8\eta'(v_{45},\tau)$, where
\begin{equation}
  \eta'(v,\tau) = -\half - \frac{1}{2\pi
  i}\frac{\theta'_1(v,\tau)}{\theta_1(v,\tau)},
\end{equation}
and where $v_{rs}\equiv v_r-v_s$. The trace of eight $S_0$ modes gives
an epsilon symbol and so $S_0^AS_0^C$ can be replaced by its
antisymmetric part, $\frac{1}{4}R_0^{ab}\gamma_{ab}^{AC}$. However,
the zero-mode trace then contains four $R_0^{ab}$ tensors, which gives
the familiar $t_8$ tensor, and so \eqref{eq:R5R5trace} evaluates to
\begin{align}
  \frac{1}{16}f(w)^8\eta'(v,\tau)\,\tr(\gamma^{a_4b_4}\gamma^{a_5b_5}\gamma^{ab})
  \,t_8^{a_1b_1a_2b_2a_3b_3ab} \equiv
  f(w)^8\eta'(v,\tau)\bar{t}_{10}^{\,a_4b_4a_5b_5a_1b_1a_2b_2a_3b_3},
\end{align}
where $\bar{t}_{10}$ is a new ten-index tensor, distinct to $t_{10}$,
and where the bar does not imply a complex conjugate. By evaluating the
trace over gamma matrices, $\bar{t}_{10}$ can be written in terms of
$t_8$ tensors as
\begin{equation} \label{eq:bart10t8}
  \bar{t}_{10}^{\,abcdefghij} = - \delta^{ad}t_8^{bcefghij}
  - \delta^{ac}t_8^{dbefghij} - \delta^{bc}t_8^{adefghij}
  - \delta^{bd}t_8^{caefghij}.
\end{equation}
There is an important relationship between the two ten-index tensors,
given in \eqref{eq:t10identity}, which relates $t_{10}$ to a sum of
ten $\bar{t}_{10}$ tensors.

So after summing over all positions of the non-zero $S$ modes and
performing an almost identical calculation for the $\tilde{S}$ modes,
the $f(w)$ terms cancel out and the $R^5\tilde{R}^5$ term is given by
\begin{align}
  & \dT \dV \prod_{r<s}(\chi_{rs})^{\frac{1}{2}k_r\cdot k_s} k_1^{b_1}
  \cdots k_5^{b_5} k_1^{d_1} \cdots k_5^{d_5} \notag\\
  & \quad \times\Big( t_{10}^{a_1b_1a_2b_2\cdots} + \sum_{r<s}
  \bar{t}_{10}^{\,a_rb_ra_sb_s\cdots}\eta'(v_{rs},\tau) \Big)
  \Big( t_{10}^{c_1d_1c_2d_2\cdots} + \sum_{r<s}
  \bar{t}_{10}^{\,c_rd_rc_sd_s\cdots}\bar\eta'(v_{rs},\tau) \Big),
\end{align}
where, for example, $\bar{t}_{10}^{\,a_2b_2a_4b_4\cdots}$ means
$\bar{t}_{10}^{\,a_2b_2a_4b_4a_1b_1a_3b_3a_5b_5}$ and $r,s$ range form
$1$ to $5$. We have suppressed the five $h_{a_rc_r}$ polarisations.

\subsubsection{Terms Containing $\partial{X}R^4\tilde{R}^5$ and
$R^5\bar\partial{X}\tilde{R}^4$} \label{sec:dXR4R5}
Consider the particular case of $\partial{X}R^4\tilde{R}^5$ where the
$\partial{X}$ originates from the first vertex operator; other cases
are practically identical. The trace over the $S$ modes gives the
usual $t_8$ tensor and the trace over the $\tilde{S}$ modes gives the
same combination of $t_{10}$ and $\bar{t}_{10}$ tensors found in the
previous section. The traces over $\alpha$ and $\tilde\alpha$ and the
$p$-integral give
\begin{equation}
  |f(w)|^{-16} \left(\frac{2}{Im\tau}\right)^5
  \prod_{r<s}(\chi_{rs})^{\frac{1}{2}k_r\cdot k_s}
  \left(-\sum_{r=2}^5 k_r^{a_1}\eta(v_{r1},\tau)\right),
\end{equation}
where $\eta(v,\tau)$ is related to the derivative of the $\chi$
function,
\begin{align} \label{eq:eta}
  \eta(v,\tau) &= \frac{1}{\pi i} \frac{\partial}{\partial v} \big(\ln
  \chi(v,\tau)\big) \notag\\
  &= \frac{\Imagi\,v}{\ImTau} + \frac{1}{2\pi i}
  \frac{\theta_1'(v,\tau)}{\theta_1(v,\tau)}.
\end{align}
Importantly, this is similar to the $\eta'(v,\tau)$ function found in
the previous section, and in particular
\begin{equation}
  \eta'(v,\tau) = -\eta(v,\tau)+\frac{\Imagi\,v}{\ImTau}-\half.
\end{equation}
So, the contribution from this $\partial{X}R^4\tilde{R}^5$ term, again
suppressing the polarisations, is
\begin{align}
  & \dT \dV \prod_{r<s}(\chi_{rs})^{\frac{1}{2}k_r\cdot k_s} k_2^{b_2}
  \cdots k_5^{b_5} k_1^{d_1} \cdots k_5^{d_5} \notag\\
  & \quad \times t_8^{a_2b_2\cdots a_5b_5}
  \Big( \sum_{r\neq 1} k_r^{a_1}\eta(v_{r1},\tau) \Big)
  \Big( t_{10}^{c_1d_1c_2nd_2\cdots} + \sum_{r<s}
  \bar{t}_{10}^{\,c_rd_rc_sd_s\cdots}\bar\eta'(v_{rs},\tau) \Big),
\end{align}
with similar expressions for the other $\partial{X}R^4\tilde{R}^5$
terms and for the $R^5\bar\partial{X}\tilde{R}^4$ terms.

\subsubsection{Terms Containing $\partial{X}\bar\partial{X}R^4\tilde{R}^4$}
Consider the case where $\partial{X}$ originates from the first vertex
and $\bar\partial{X}$ from the second. The $S$ and $\tilde{S}$ traces
give a pair of $t_8$ tensors and the $\alpha$ and $\tilde\alpha$
traces lead to
\begin{equation}
|f(w)|^{-16} \left(\frac{2}{Im\tau}\right)^5
\prod_{r<s}(\chi_{rs})^{\frac{1}{2}k_r\cdot k_s}
\left(\sum_{r\neq 1} k_r^{a_1}\eta(v_{r1},\tau)\sum_{s\neq 2} k_s^{c_2}\bar\eta(v_{s2},\tau)
- 2\delta^{a_1c_2}\hat\Omega(v_{12},\tau) \right),
\end{equation}
where $\hat\Omega(v,\tau)$ is given by
\begin{align} \label{eq:OmegaHat}
  \hat\Omega(v,\tau) &= \frac{1}{\pi^2} \frac{\partial}{\partial v}
  \frac{\partial}{\partial \bar v} \big( \ln \chi(v,\tau) \big) \notag\\
  &= -\frac{1}{2\pi\ImTau} + \frac{1}{2\pi}\delta^2(v).
\end{align}
It is unclear whether the $\delta^2(v)$ factor should be included since
it is not seen by performing the integral over $p$. However, it makes
no difference for the five-point amplitude since, after performing the
integral over $v$ and analytically continuing to the region where
$k_r\cdot k_s>0$, it leads to a vanishing $\chi(0,\tau)^{\half k_r\cdot k_s}$
factor. So this particular $\partial{X}\bar\partial{X}R^4\tilde{R}^4$
term gives
\begin{align}
  & \dT \dV \prod_{r<s}(\chi_{rs})^{\frac{1}{2}k_r\cdot k_s} k_2^{b_2}
  \cdots k_5^{b_5} k_1^{d_1}k_3^{d_3} \cdots k_5^{d_5} \notag\\
  &\quad\times \left(\sum_{r=2}^5
  k_r^{a_1}\eta(v_{r1},\tau)\sum_{s=1,s\neq 2}^5
  k_s^{c_2}\bar\eta(v_{s2},\tau)
  - 2\delta^{a_1c_2}\hat\Omega(v_{12},\tau)
  \right)t_8^{a_2b_2a_3b_3\cdots}t_8^{c_1d_1c_3d_3\cdots},
\end{align}
where the $\eta$ and $\bar\eta$ sums can be understood as a product of
the left- and right-moving modes; it is the $\hat\Omega(v,\tau)$
function, which originates in the $p$-integral, which contains the
mixing between left- and right-movers, and distinguishes the amplitude
from simply the product of two open-string amplitudes.

In the case that the $\partial X$ and $\bar\partial X$ come from the
same vertex operator, the $2\delta^{a_rc_r}\hat\Omega(v,\tau)$ term
does not contribute to the amplitude since $\delta^{a_rc_r}h_{a_rc_r}$
vanishes for a traceless graviton.

\subsection{The Overall Amplitude}
Before writing the overall amplitude, we simplify it using various
relationships between the tensors $t_{10},\,\bar{t}_{10}$ and
$t_8$. First, using identity \eqref{eq:t10identity}, we can rewrite
\begin{equation}
  t_{10}^{a_1b_1a_2b_2\cdots} + \sum_{r<s}
  \bar{t}_{10}^{\,a_rb_ra_sb_s\cdots}\eta'(v_{rs},\tau) =
  \sum_{r<s} \bar{t}_{10}^{\,a_rb_ra_sb_s\cdots}\left(
  \eta'(v_{rs},\tau)+{\textstyle\half} \right).
\end{equation}
Then, since identity \eqref{eq:bart10identity} implies $\sum_{r<s}
v_{rs}\bar{t}_{10}^{\,a_rb_ra_sb_s\cdots} = 0$, which in
turn implies
\begin{equation}
  \sum_{r<s} (\Imagi\,v_{rs})
  \bar{t}_{10}^{\,a_rb_ra_sb_s\cdots}=0,
\end{equation}
we are free to add a term linear in $\Imagi\,v$ to
$\left(\eta'(v_{rs},\tau)+\half\right)$. In particular, if we subtract
$\frac{\Imagi\,v}{\ImTau}$ then we recognise the resulting term as
$\eta(v_{rs},\tau)$, the same function found in section
\ref{sec:dXR4R5}. So we can eliminate the tensor $t_{10}$ in favour of
$\bar{t}_{10}$ and eliminate the function $\eta'(v,\tau)$ in favour of
$\eta(v,\tau)$.

Then we rewrite the $\bar{t}_{10}$ terms in terms of $t_8$ tensors
using \eqref{eq:bart10t8}, leaving the final expression for the
amplitude entirely in terms of $t_8$ tensors,
\begin{align} \label{eq:amp}
  A_{5h,t_8t_8} &=
  h^1_{a_1c_1}h^2_{a_2c_2}h^3_{a_3c_3}h^4_{a_4c_4}h^5_{a_5c_5}
  \dT \dV \chiprod \notag\\
  & \qquad\times \left( \sum_{r<s}\eta(v_{rs},\tau)A_{rs}
  \sum_{r<s}\bar\eta(v_{rs},\tau)\bar{A}_{rs}
  + \sum_{r<s}\hat\Omega(v_{rs},\tau)B_{rs} \right),
\end{align}
where
\begin{align} \label{eq:A12}
  A_{12} &= k_1^{a_2}(k_1+k_2)^{b}k_3^{b_3}k_4^{b_4}k_5^{b_5}t_8^{a_1ba_3b_3a_4b_4a_5b_5} \notag\\
  & \quad -k_2^{a_1}(k_1+k_2)^{b}k_3^{b_3}k_4^{b_4}k_5^{b_5}t_8^{a_2ba_3b_3a_4b_4a_5b_5} \notag\\
  & \quad -\delta^{a_1a_2}k_1^{b_1}k_2^{b_2}k_3^{b_3}k_4^{b_4}k_5^{b_5}t_8^{b_1b_2a_3b_3a_4b_4a_5b_5} \notag\\
  & \quad -k_1\cdot k_2\,k_3^{b_3}k_4^{b_4}k_5^{b_5}t_8^{a_1a_2a_3b_3a_4b_4a_5b_5}
\end{align}
and
\begin{align} \label{eq:B12}
  B_{12} = -4\,\delta^{a_1c_2}
  k_2^{b_2}k_3^{b_3}k_4^{b_4}k_5^{b_5} k_1^{d_1}k_3^{d_3}k_4^{d_4}k_5^{d_5}
  t_8^{a_2b_2a_3b_3a_4b_4a_5b_5}t_8^{c_1d_1c_3d_3c_4d_4c_5d_5}.
\end{align}
$\bar{A}_{12}$ is the same as $A_{12}$ but with $c_r$ replacing
$a_r$. The other $A_{rs}$ and $B_{rs}$ are similar but with the
relevant permutations of the momenta and polarisation indices. For
brevity we have suppressed the indices on $A_{rs}$, $\bar{A}_{rs}$ and
$B_{rs}$.

With the amplitude in this form, it is particularly easy to
demonstrate modular invariance, which in turn implies
finiteness. Consider a general $SL(2,\mathbb{Z})$ transformation given
by
\begin{equation}
  v_r \to \frac{v_r}{c\tau+d}, \qquad
  \tau \to \frac{a\tau+b}{c\tau+d},
\end{equation}
where $a,b,c,d\in\mathbb{Z}$ and $ad-bc=1$. Then it is easy to show
that
\begin{equation}
  d^2v \to \frac{d^2v}{|c\tau+d|^2}, \qquad
  d^2\tau \to \frac{d^2\tau}{|c\tau+d|^4}, \qquad
  \ImTau \to \frac{\ImTau}{|c\tau+d|^2},
\end{equation}
and, by using the transformation properties of $\theta_1(v,\tau)$,
that
\begin{align}
  \chi(v,\tau) \to \frac{\chi(v,\tau)}{|c\tau+d|}, &\qquad
  \eta(v,\tau) \to (c\tau+d)\eta(v,\tau), \notag\\
  \bar\eta(v,\tau) \to (c\bar\tau+d)\bar\eta(v,\tau), &\qquad
  \hat\Omega(v,\tau) \to |c\tau+d|^2\hat\Omega(v,\tau).
\end{align}
By using $\sum_{r<s} k_r\cdot k_s=0$, which follows from momentum
conservation, it is then simple to check that \eqref{eq:amp} is
modular invariant.

Gauge invariance is guaranteed from the gauge invariance of
\eqref{eq:5gravAmp}, but it is reassuring to confirm this explicitly
by replacing, say, $h_1^{ac}$ by $k_1^a\zeta^c+k_1^c\zeta^a$ where
$k_1\cdot\zeta=0$. In fact, it is sufficient to replace $h_1^{ac}$ by
$k_1^a\zeta^c$. Then gauge invariance is easily shown using the
antisymmetry of $t_8$, the vanishing of
\begin{equation}
  \dT \dV \, \frac{\partial}{\partial v_1}\left(
  \prod_{r<s}\chi_{rs}^{\half k_r \cdot k_s}
  \frac{\partial}{\partial\bar{v}}\ln\chi(v_{12},\tau) \right)
\end{equation}
when integrated over a surface with no boundary, and the identity
\eqref{eq:bart10identity}.

Bose symmetry, \ie\ symmetry under
$(h_r,k_r,v_r)\leftrightarrow(h_s,k_s,v_s)$, is a trivial consequence
of the symmetries of $t_8$ and the fact that under $v\to-v$,
\begin{equation}
  \eta(v,\tau)\to-\eta(v,\tau), \qquad
  \hat\Omega(v,\tau)\to\hat\Omega(v,\tau).
\end{equation}

\subsection{Convergence Issues} \label{sec:5gravConv}
As explained in more detail in chapter \ref{sec:5gravExp}, there are
now four different integrals over $v_r$ and $\tau$ which must be
performed. We concentrate here on
\begin{equation} 
  I_{12} = \dT \dV \chiprod |\eta_{12}|^2,
\end{equation}
where $\eta_{12}\equiv\eta(v_{12},\tau)$, since it leads, along with
similar integrals, to the most stringent restrictions on the
convergence. As with the four-graviton amplitude, there are two corners
of the integration region which can lead to singularities. The first
is when $v_1\to v_2$ which can be examined by writing
$v_{12}=|v|e^{i\theta}$ and integrating over a small region
near$|v|=0$. Since for small $v$,
\begin{equation}
  \chi(v,\tau) \sim 2\pi|v|, \qquad
  \eta(v,\tau) \sim -\frac{i}{2\pi v},
\end{equation}
we find, suppressing all integrals except $v_{12}$,
\begin{equation} \label{eq:Ipoles}
  I_{12} \sim \int d^2v |v|^{-\quart s} \frac{1}{|v|^2} \sim
  \int d|v||v|^{-\quart s-1},
\end{equation}
where $s\equiv -2k_1\cdot k_2$, which diverges unless $s<0$. Contrast
this with the equivalent condition for the four-graviton amplitude,
$s<8$, with the difference being due to the extra $|\eta|^2$ factor
for five gravitons.

The second potential singularity is due to the region
$\ImTau\to\infty$ when the torus degenerates to a thin wire. As with
the four-graviton amplitude, we study this by splitting the integral
into twelve parts depending on the $v_r$ orderings and rewriting
$v_{rs}=x+\tau y$, where $x$ and $y$ are real variables with $0\leq
x,y<1$. Since the integration region no longer contains an $\ImTau$
factor, we can study the $\ImTau\to\infty$ behaviour just from
studying the integrand. In terms of $x$ and $y$, it is easy to show
that for large $\ImTau$,
\begin{equation*}
  \chi(v,\tau) \sim e^{\pi\Imagi\,\tau\cdot y(1-y)},
\end{equation*}
and that $\eta(v,\tau)$ tends to a constant. Then since $\chi^{-\quart
s}$ appears in the integrand, it can be shown that the $\tau$-integral
will only converge if $s\leq 0$. This is the same restriction found in
the four-graviton case. If all twelve regions are combined then this
constraint is extended to $s\leq 0$, $t\leq 0$, $u\leq 0$, $v\leq 0$,
$w\leq 0$ and $x\leq 0$.

It is not possible to simultaneously satisfy both the $v_{rs}\to 0$
and $\ImTau\to\infty$ constraints and so the integral does not
converge anywhere. Extending $s$ to complex values does not help and
the integral diverges even for purely imaginary values. Na\"ively this
is a disaster since the amplitude is nowhere finite leaving no hope
for an analytic continuation. However, although the full integral
converges nowhere, this is not true for the separate regions and so
the resolution is to separately analytically continue each of the
twelve pieces. After doing so, the full amplitude will then contain
all the correct massless poles, massive poles and branch-cut
singularities required by unitarity \cite{Montag:1992dm}.

%%%%%%%%%%%%%%%%%%%%%%%%%%%%%%%%%%%%%%%%%%%%%%%%%%%%%%%%%%%%%%%%%%%%%%%%%%%%%%%%%%%%%%%%%%%%%%%%%%%%%%%%%%%%%%
\section{Expansion of the Five-Graviton Amplitude} \label{sec:5gravExp}
%%%%%%%%%%%%%%%%%%%%%%%%%%%%%%%%%%%%%%%%%%%%%%%%%%%%%%%%%%%%%%%%%%%%%%%%%%%%%%%%%%%%%%%%%%%%%%%%%%%%%%%%%%%%%%
Unlike the four-graviton amplitude which only involves the single
integral \eqref{eq:4gravI}, the five-graviton amplitude contains four
separate integrals, which we denote by
\begin{align} \label{eq:ints}
  I_{rs} &= \dT \dVdash \chiproddash |\eta_{rs}|^2, \notag \\
  J_{rs|rt} &= \dT \dVdash \chiproddash \eta_{rs}\bar\eta_{rt}, \notag \\
  J'_{rs|tu} &= \dT \dVdash \chiproddash \eta_{rs}\bar\eta_{tu}, \notag \\
  K &= \dT \dVdash \chiproddash \hat\Omega_{rs},
\end{align}
where $r,\,s,\,t,\,u$ are all different and where
$\eta_{rs}\equiv\eta(v_{rs},\tau)$ and similarly for $\chi_{rs}$,
$\bar\eta_{rs}$ and $\hat\Omega_{rs}$. Since $\hat\Omega_{rs}$ is
independent of $v_{rs}$, we write $K$ without any subscripts. We want
to study the effective action up to and including $D^4R^5$
terms. Since the $I$, $J$ and $J'$ integrals are multiplied by ten
momenta, this implies they need to be expanded up to order
$\alpha'^2$; whereas since the $K$ integral is multiplied by only
eight momenta, the expansion to order $\alpha'^3$ is required. These
expansions are obtained using similar techniques to those used for the
four-graviton case. It is important to remove the massless pole in
$I_{rs}$ before attempting to expand about zero momenta. Threshold
branch cuts will again be removed by only integrating $\tau$ over the
region of the fundamental domain with $\ImTau\le L$.

\subsection{Five-particle Mandelstam Variables}
The integrals \eqref{eq:ints} are parameterized by products of the
momenta, $k_r\cdot k_s$. However, due to momentum conservation, these
are not independent quantities. In order to find the expansions of
\eqref{eq:ints}, it is important to use an independent set of such
products, the Mandelstam variables, the equivalent of $s$ and $t$ in
the four-graviton case. Consider the ten variables $k_r\cdot k_s$ with
$r<s$. Momentum conservation can be used to eliminate, say, the four
involving $k_5$. The remaining six, however, are still not independent
since we can still impose $k_5^2=(-k_1-k_2-k_3-k_4)^2=0$, and we use
this to eliminate $k_3\cdot k_4$. We choose to label the remaining
five independent variables by
\begin{equation} \label{eq:5gravMand1}
  s = -2k_1\cdot k_2,\quad t = -2k_1\cdot k_3,\quad u = -2k_1\cdot k_4,
 \quad v = -2k_2\cdot k_3,\quad w = -2k_2\cdot k_4,
\end{equation}
with a sixth non-independent variable given by
\begin{equation}
  x = -2k_3\cdot k_4,
\end{equation}
where $s+t+u+v+w+x=0$. The remaining invariants are then given by
\begin{align} \label{eq:5gravMand2}
  -2k_1\cdot k_5 &= -s-t-u, \qquad -2k_2\cdot k_5 = -s-v-w, \notag\\
  -2k_3\cdot k_5 &= s+u+w, \qquad\:\: -2k_4\cdot k_5 = s+t+v.
\end{align}

Integrals \eqref{eq:ints} have, by construction, certain symmetries in
some of the $k_r\cdot k_s$. For example, $K$ is manifestly symmetric
in all the $k_r\cdot k_s$. For four gravitons this symmetry manifests
itself as a symmetry in $s$, $t$ and $u$. However, for five gravitons
this is no longer the case: the asymmetry between
\eqref{eq:5gravMand1} and \eqref{eq:5gravMand2} means that any
symmetry in $k_r\cdot k_s$ is hidden when written in the $s$, $t$,
$u$, $v$, $w$, $x$ variables. This means, at least for $J$ and $J'$,
that we are unable to guess an ansatz, akin to \eqref{eq:4gravExp},
for the expansion in terms of $k_r\cdot k_s$. Instead, the best we can
do is to find the most general expansion in terms of $s$, $t$, $u$,
$v$, $w$ and reverse engineer to find a, hopefully unique, expansion
in terms of $k_r\cdot k_s$.

\subsection{Massless Poles in Integral $I$} \label{sec:Ipoles}
\begin{figure}
\begin{center}
\includegraphics[scale=0.6]{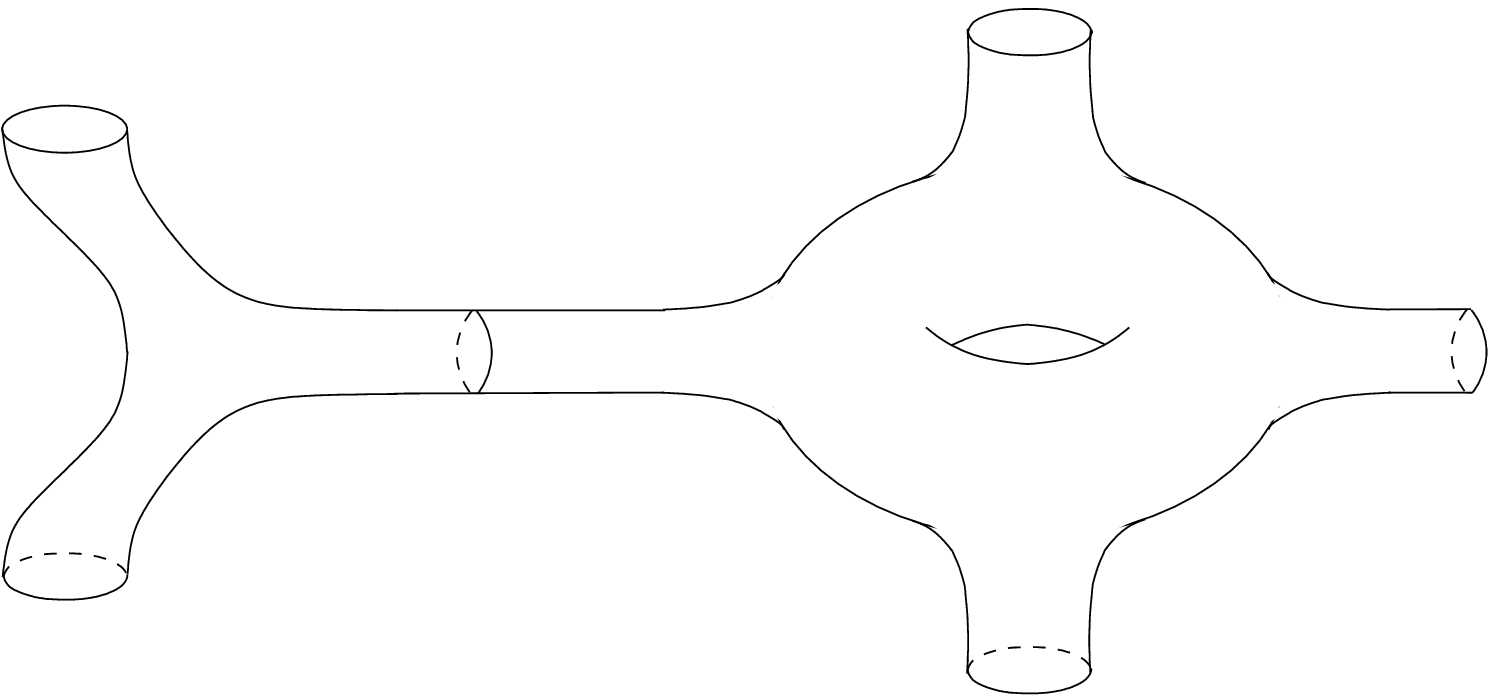}
\caption{The origin of the massless poles in the five-graviton
amplitude.}
\label{fig:5gravPole}
\end{center}
\end{figure}

Unlike the four-graviton amplitude, which contains no massless poles
due to the vanishing of the three-graviton one-loop amplitude, the
five-graviton version does contain such poles corresponding to the
string diagram shown in figure \ref{fig:5gravPole}. These massless
poles originate from the integral $I_{rs}$ and can be studied by
considering the limit as two vertex operators approach each other on
the world-sheet.

By extending the analysis in section \ref{sec:5gravConv}, the limit as
$v_2\to v_1$ identifies the massless pole in $s$ as
\begin{equation} \label{eq:Ipoles2}
  I_{12} \sim -\frac{1}{s} \cdot \frac{2}{\pi} \dT \int_{\cal T}\prod_{r=2}^4 d^2v_r
  {\prod_{\substack{r<s \\ 1 \to 2}}}'(\chi_{rs})^{\half k_r\cdot k_s},
\end{equation}
where the prime indicates that $(r,s)=(1,2)$ is not to be included,
and $1 \to 2$ means that $v_1$ is to be replaced by $v_2$ everywhere
within the product. This product can be rewritten without the prime as
$\prod_{r<s}(\chi_{rs})^{\half k_r'\cdot k_s'}$, where $r,\,s$ now run
only from $1$ to $4$ and where
\begin{equation}
  k_1'=k_1+k_2,\qquad k_2'=k_3,\qquad k_3'=k_4,\qquad k_4'=k_5.
\end{equation}
It is now easy to recognise the residue of the pole as the product of
a tree-level three-graviton vertex and a four-graviton one-loop
amplitude, as required by unitarity. The poles for other values of $r$
and $s$ work in an identical manner.

\subsection{Expansion of Integral $K$}
First we expand integral $K$ since this is most similar to the
integral in the four-graviton case. Ignoring the irrelevant delta
function in \eqref{eq:OmegaHat},
\begin{equation}
  K = -\frac{1}{2\pi} \int_{\cal F}\!\frac{d^2\tau}{(\ImTau)^6} \dVdash
  \chiproddash,
\end{equation}
where the product over $\chi$'s can be written in terms of independent
Mandelstam variables as
\begin{equation}
  \left(\frac{\chi_{12}\chi_{35}\chi_{45}}{\chi_{15}\chi_{25}\chi_{34}}\right)^{-\quart s}
  \left(\frac{\chi_{13}\chi_{45}}{\chi_{15}\chi_{34}}\right)^{-\quart t}
  \left(\frac{\chi_{14}\chi_{35}}{\chi_{15}\chi_{34}}\right)^{-\quart u}
  \left(\frac{\chi_{23}\chi_{45}}{\chi_{25}\chi_{34}}\right)^{-\quart v}
  \left(\frac{\chi_{24}\chi_{35}}{\chi_{25}\chi_{34}}\right)^{-\quart w}.
\end{equation}
By studying $v_r\to v_s$, it is easy to show that there are no
massless poles.

Since none of the $v_r$ play a privileged r\^ole in the integrand of
$K$, the low-energy expansion will be symmetric in the variables
$k_r\cdot k_s$ and so we require the most general symmetric
expressions at each order. At order $\alpha'^2$ there are three
potential candidates, which are in fact all proportional to each
other,
\begin{align}
  (k_1\cdot k_2)^2 + (k_1\cdot k_3)^2 + \cdots + (k_4\cdot k_5)^2 &\equiv X, \notag\\
  (k_1\cdot k_2)(k_1\cdot k_3) + \cdots
  + (k_3\cdot k_5)(k_4\cdot k_5) &= -X, \notag\\
  (k_1\cdot k_2)(k_3\cdot k_4) + \cdots
  + (k_2\cdot k_5)(k_3\cdot k_4) &= {\textstyle\half} X.
\end{align}
At order $\alpha'^3$ there are many more symmetric expressions, but
again it can be shown that they are all proportional to each other, so
that we are free to choose $\sum_{r<s}(k_r\cdot k_s)^3$ as the only
independent combination. Then the most general expansion for $K$ up to
order $\alpha'^3$ is given by
\begin{align} \label{eq:Kexp}
  K &=  a + b\sum_{r<s} k_r\cdot k_s + c\sum_{r<s} (k_r\cdot k_s)^2 +
  d\sum_{r<s} (k_r\cdot k_s)^3 \notag\\
  &= a + \frac{c}{4}( 6s^2+4t^2+6st+\cdots ) + \frac{d}{8}( 6t^2u+\cdots ),
\end{align}
where the second term has been dropped since it vanishes using
momentum conservation. Threshold terms have been ignored since these
will be removed by imposing $\ImTau<L$ on the fundamental domain.

The value of $a$ is found simply by setting $s=t=u=v=w=0$, giving
\begin{equation}
   a = -\frac{1}{2\pi} \int_{\cal F}\!\frac{d^2\tau}{(\ImTau)^6} \dVdash
     = -\frac{1}{2\pi} \int_{\cal F}\!\frac{d^2\tau}{(\ImTau)^2}
     = -\frac{1}{6}.
\end{equation}

Using \eqref{eq:Kexp}, the value of $c$ can be found in many ways,
all, of course, giving the same result. For example,
\begin{align}
  c &= \lim_{s,t,\ldots\to 0} \left( \half \frac{\partial^2 K}{\partial
  t^2} \right)\bigg|_{\rm const} \notag\\
  &= -\frac{1}{2^6\pi} \int_{\cal F}\!\frac{d^2\tau}{(\ImTau)^6} \dVdash \,
  (\ln\chi_{13}+\ln\chi_{45}-\ln\chi_{15}-\ln\chi_{34})^2\,\bigg|_{\rm const}.
\end{align}
After expanding the bracket only the square terms will give non-zero
contributions. For example, $\dVdash \ln\chi_{13} \ln\chi_{45}$
vanishes since the integral of a single $\ln\chi$ is zero. By changing
variables and performing three of the $v_{r'}$-integrals,
\begin{equation}
  c = -\frac{1}{2^4\pi} \int_{\cal F}\!\frac{d^2\tau}{(\ImTau)^3}
  \int_{\cal T} d^2v (\ln\chi)^2\,\bigg|_{\rm const}
  \equiv -\frac{1}{2^4\pi} \Theta_1|_{\rm const},
\end{equation}
where the last part defines $\Theta_1$. The integral $\Theta_1$ is
exactly the same as that encountered in section \ref{sec:4gravB} and
when the remaining $v$-integral is performed we again find the Epstein
zeta function $Z_2(\tau,\bar\tau)$. Following the same analysis, the
$\tau$-integral over the $\ImTau<L$ region of the fundamental domain
gives
\begin{equation}
  c_L = -\frac{\pi}{2^2\cdot 6!}L + \mathcal{O}(L^{-2})
  \quad\Rightarrow\quad c = 0,
\end{equation}
where the final part follows since $L$-dependent parts of $c_L$ must
cancel with the same integral over the upper part of the fundamental
domain.

Finding $d$ involves a similar calculation. For example,
\begin{align}
  d &= \frac{2}{3} \lim_{s,t,\ldots\to 0} \frac{\partial^3 K}{\partial
  t^2 \partial u}\bigg|_{\rm const} \notag\\
  &= \frac{1}{2^6\cdot 3\pi} \int_{\cal F}\!\frac{d^2\tau}{(\ImTau)^6} \dVdash \,
  (\ln\chi_{13}+\ln\chi_{45}-\ln\chi_{15}-\ln\chi_{34})^2 \notag\\
  &\qquad\qquad\times (\ln\chi_{14}+\ln\chi_{35}-\ln\chi_{15}-\ln\chi_{34})
  \,\Big|_{\rm const}.
\end{align}
There are only two non-vanishing contributions: eight terms containing
$\ln\chi_{rs}\ln\chi_{st}\ln\chi_{tr}$ and two terms containing
$(\ln\chi_{rs})^3$. After performing the $v$-integrals for the first
kind, the contribution to $d$ is given by
\begin{equation}
  -\frac{1}{2^3\cdot 3\pi} \int_{\cal F}\!\frac{d^2\tau}{(\ImTau)^4}
  \int_{\cal T} d^2v_1 d^2v_2 \left( \ln\chi_1\ln\chi_2\ln\chi_{1+2} \right) \equiv
  -\frac{1}{2^3\cdot 3\pi} \Theta_2,
\end{equation}
which defines $\Theta_2$ and where $\ln\chi_{1+2}$ means
$\ln\chi(v_1+v_2,\tau)$. The same integral was found in section
\ref{sec:4gravC} where it was shown to involve the zeta function
$Z_3(\tau,\bar\tau)$. As there, the $\tau$-integral can be converted
to an integral over the boundary of the restricted fundamental domain
which again leads to $L$-dependent terms, but no constant piece. So
$d$ is given entirely by the second type of term,
\begin{equation}
  d = -\frac{1}{2^5\cdot 3\pi} \int_{\cal F}\!\frac{d^2\tau}{(\ImTau)^3}
  \int_{\cal T} d^2v (\ln\chi)^3\,\bigg|_{\rm const}
  \equiv -\frac{1}{2^5\cdot 3\pi} \Theta_3|_{\rm const}.
\end{equation}
The integral $\Theta_3$ was also encountered in section
\ref{sec:4gravC} and, using the same `unfolding procedure' as in
\cite{Green:1999pv}\footnote{Note that the definition of $\ln\chi$ used
here differs from that in \cite{Green:1999pv} by a factor of $-2$:
$[\ln\chi]_{\rm here}=-2 [\ln\chi]_{\rm there}$.}, it is easy to show
that
\begin{equation}
  \Theta_3|_{\rm const} = -\frac{\zeta(2)\zeta(3)}{4\pi}.
\end{equation}

So, up to order $\alpha'^3$ and ignoring threshold corrections, the
expansion of $K$ is given by
\begin{equation} \label{eq:FinalKexp}
  K = -\frac{1}{6} + \frac{\zeta(3)\alpha'^3}{2^5\cdot 3^2} \sum_{r<s}
  (k_r\cdot k_s)^3 + \mathcal{O}(\alpha'^4),
\end{equation}
where $\alpha'$ has been reinstated using $2\alpha'=1$.

\subsection{Expansion of Integral $J$} \label{sec:Jexp}
Now consider the integral $J_{rs|rt}$, which only needs to be expanded
up to order $\alpha'^2$. For concreteness consider $J_{12|13}$. Then,
using \eqref{eq:eta} and changing variables so that $v_{12} \to v_1$,
$v_{13} \to v_2$,
\begin{align} \label{eq:Jexp}
  J_{12|13} &= \frac{1}{\pi^2}\dT \dVdash \,
  \frac{\partial}{\partial v}(\ln\chi_1) \frac{\partial}{\partial
  \bar v}(\ln\chi_2) \notag\\
  & \qquad\times
    \left(\frac{\ln\chi_1\ln\chi_3\ln\chi_4}{\ln\chi_{2+3}\ln\chi_{-1+2+3}\ln\chi_{3-4}}\right)^{-\quart s}
    \left(\frac{\ln\chi_2\ln\chi_4}{\ln\chi_{2+3}\ln\chi_{3-4}}\right)^{-\quart t} \notag\\
  & \qquad\times
    \left(\frac{\ln\chi_{2+3-4}\ln\chi_3}{\ln\chi_{2+3}\ln\chi_{3-4}}\right)^{-\quart u}
    \left(\frac{\ln\chi_{-1+2}\ln\chi_4}{\ln\chi_{-1+2+3}\ln\chi_{3-4}}\right)^{-\quart v}
    \left(\frac{\ln\chi_{-1+2+3-4}\ln\chi_3}{\ln\chi_{-1+2+3}\ln\chi_{3-4}}\right)^{-\quart w},
\end{align}
where, due to the periodicity of the integrand, there is no change to
the $v_r$ integration region. The absence of massless poles can again
be shown by studying the $v_{rs}\to 0$ limit.

Unlike $K$, $v_{rs}$ and $v_{rt}$ now play a privileged r\^ole in the
definition of $J_{rs|rt}$ and so the expansion is expected to have
less symmetry. However, $J_{12|13}$ should still be symmetric under
$k_4\leftrightarrow k_5$ and, assuming it is real, under
$k_2\leftrightarrow k_3$. This symmetry is manifest when the expansion
is written in terms of $k_r\cdot k_s$, which is also the form required
for comparing with the effective action. However, in practice we
determine the expansion in terms of the independent Mandelstam
variables $s$, $t$, $u$, $v$ and $w$, for which any symmetry is lost,
and then deduce the expansion in terms of $k_r\cdot k_s$. So we must
consider the most general form for the expansion up to order
$\alpha'^2$,
\begin{align} \label{eq:Jexp2}
  J_{12|13} &=\: a \notag\\
    &\quad + b_1s + b_2t + b_3u + b_4v + b_5w \notag\\
    &\quad + c_1s^2 + c_2t^2 + c_3u^2 + c_4v^2 + c_5w^2 \notag\\
    &\quad + d_1st + d_2su + d_3sv + d_4sw + d_5tu + d_6tv +d_7tw + d_8uv
       + d_9uw + d_{10}vw,
\end{align}
where, as usual, we have ignored non-analytic terms.

The constant $a$ is easily found by setting all the Mandelstam
variables to zero, leaving an expression involving the integral $\int_{\cal T}
d^2v \frac{\partial}{\partial v}(\ln\chi)$. Although the
integrand is infinite at $v=0$, the integral itself is finite, as can
be seen by writing $d^2v$ as $|v|d|v|d\theta$ and, in fact, vanishes
due to the antisymmetry of $\partial_v\ln\chi$ under $v\to-v$.

The coefficients $b_i$ are determined by considering single
derivatives of \eqref{eq:Jexp} with respective to some Mandelstam
variable. For example,
\begin{align}
  b_1 &= \lim_{s,t,\ldots\to 0}\frac{\partial J}{\partial s}\bigg|_{\rm const} \notag\\
      &= -\frac{1}{4\pi^2} \dT\dVdash\, \frac{\partial}{\partial v}
         (\ln\chi_1) \frac{\partial}{\partial \bar v}
         (\ln\chi_2) \notag\\
      &\qquad\quad\times \big(\ln\chi_1+\ln\chi_3+\ln\chi_4
       -\ln\chi_{2+3}-\ln\chi_{-1+2+3}-\ln\chi_{3-4} \big)\,\Big|_{\rm const},
\end{align}
where all the terms vanish since they all contain $\int_{\cal T}
d^2v\,\partial_v\ln\chi$ factors. (To see this for the fifth term it
is necessary to change variables so that $-v_1+v_2+v_3 \to v_3$.)
Similarly $b_2$, $b_3$ and $b_5$ all vanish. However, $b_4$ is
potentially non-zero,
\begin{equation}
  b_4 = \frac{1}{4\pi^2} \int_{\cal F}\!\frac{d^2\tau}{(\ImTau)^3}
         \int_{\cal T} d^2v_1 d^2v_2\, \frac{\partial}{\partial v}
         (\ln\chi_1) \frac{\partial}{\partial \bar v} (\ln\chi_2)
         \ln\chi_{1+2}\,\bigg|_{\rm const}
      \equiv \frac{1}{4\pi^2} \Theta_4|_{\rm const}.
\end{equation}
It is worth noting that, despite appearances, $\Theta_4$ is still
modular invariant since the constant term added to $\ln\chi_{1+2}$ under a
modular transformation vanishes by the antisymmetry of
$\partial_v\ln\chi$. After performing the $v$-integrals, the same zeta
function $Z_2(\tau,\bar\tau)$ is found as at order $\alpha'^2$ in the
expansion of $K$, despite originating from a different integral. In
fact, $\Theta_4=\half\pi\Theta_1$. Since $\Theta_1$ only leads to
$L$-dependent terms with no constant part, we can conclude that
$b_4=0$.

The $c_i$ and $d_i$ coefficients can be found in a similar manner
giving
\begin{equation}
  c_1 = \frac{1}{16\pi^2}\Theta_5|_{\rm const}, \quad c_2 = 0, \quad c_3=0,
  \quad c_4= -\frac{1}{32\pi^2}\Theta_6|_{\rm const},
  \quad c_5 = \frac{1}{8\pi^2}\Theta_5|_{\rm const},
\end{equation}
and
\begin{align}
  d_1 &= 0, & d_2 &= \frac{1}{16\pi^2}\Theta_5|_{\rm const},
  & d_3 &= \frac{1}{16\pi^2}(\Theta_5|_{\rm const}-\Theta_7|_{\rm const}), \notag\\
  d_4 &= \frac{3}{16\pi^2}\Theta_5|_{\rm const}, & d_5 &= 0,
  & d_6 &= -\frac{1}{16\pi^2}\Theta_8|_{\rm const}, \notag\\
  d_7 &= \frac{1}{16\pi^2}\Theta_5|_{\rm const}, & d_8 &= \frac{1}{16\pi^2}\Theta_5|_{\rm const},
  & d_9 &= \frac{1}{8\pi^2}\Theta_5|_{\rm const}, \notag\\
  d_{10} &= \frac{1}{8\pi^2}\Theta_5|_{\rm const},
\end{align}
where
\begin{align}
  \Theta_5 &\equiv \int_{\cal F}\!\frac{d^2\tau}{(\ImTau)^4} \int_{\cal T} d^2v_1d^2v_2d^2v_3\,
              \frac{\partial}{\partial v}(\ln\chi_1)
              \frac{\partial}{\partial \bar v}(\ln\chi_2)
              \ln\chi_{1+2+3}\ln\chi_3, \notag\\
  \Theta_6 &\equiv \int_{\cal F}\!\frac{d^2\tau}{(\ImTau)^3} \int_{\cal T} d^2v_1d^2v_2\,
              \frac{\partial}{\partial v}(\ln\chi_1)
              \frac{\partial}{\partial \bar v}(\ln\chi_2)
              (\ln\chi_{1+2})^2, \notag\\
  \Theta_7 &\equiv \int_{\cal F}\!\frac{d^2\tau}{(\ImTau)^3} \int_{\cal T} d^2v_1d^2v_2\,
              \frac{\partial}{\partial v}(\ln\chi_1)
              \frac{\partial}{\partial \bar v}(\ln\chi_2)
              \ln\chi_1\ln\chi_{1+2}, \notag\\
  \Theta_8 &\equiv \int_{\cal F}\!\frac{d^2\tau}{(\ImTau)^3} \int_{\cal T} d^2v_1d^2v_2\,
              \frac{\partial}{\partial v}(\ln\chi_1)
              \frac{\partial}{\partial \bar v}(\ln\chi_2)
              \ln\chi_{2}\ln\chi_{1+2}.
\end{align}
Despite involving different integrands, performing the $v_r$-integrals
shows that $\Theta_5$ and $\Theta_6$ involve the same modular
functions encountered in calculating $d$ in the expansion of $K$. In
fact, $\Theta_5 = \frac{\pi}{2}\Theta_2$ and $\Theta_6 =
\frac{\pi}{2}\Theta_3$ and so, after performing the $\tau$-integrals,
we find
\begin{equation} \label{eq:Theta56const}
  \Theta_5|_{\rm const} = 0, \qquad \Theta_6|_{\rm const} =
  -\frac{\zeta(2)\zeta(3)}{8}.
\end{equation}
However, $\Theta_7$ and $\Theta_8$ do lead to new expressions,
\begin{align}
  \Theta_7 &= -\frac{1}{16\pi^2} \int_{\cal F}\!\frac{d^2\tau}{(\ImTau)^2}
              \sum_{(m,n),(p,q),(r,s)\neq(0,0)}
              \frac{(\ImTau)^3\,\delta_{m+r-p}\delta_{n+s-q}}{(m\tau+n)(p\bar\tau+q)|r\tau+s|^2|p\tau+q|^2},
              \notag\\
  \Theta_8 &= -\frac{1}{16\pi^2} \int_{\cal F}\!\frac{d^2\tau}{(\ImTau)^2}
              \sum_{(m,n),(p,q),(r,s)\neq(0,0)}
              \frac{(\ImTau)^3\,\delta_{m+r-p}\delta_{n+s-q}}{(m\tau+n)(p\bar\tau+q)|r\tau+s|^2|m\tau+n|^2}.
\end{align}
As with $\Theta_3$ (and $\Theta_6$), there is no obvious way to write
the integrands as total derivatives. Perhaps an `unfolding procedure'
could be used as in \cite{Green:1999pv}. However, it is not necessary
to explicitly evaluate them since their constant parts will be
inferred as follows. The same integrals will appear in the pole
terms of the expansion of integral $I_{rs}$. However, these poles are
completely fixed by unitarity and this will allow $\Theta_7$ and
$\Theta_8$ to be uniquely determined. For completeness we state here
that we will find
\begin{equation} \label{eq:Theta78}
  \Theta_7|_{\rm const} = \Theta_8|_{\rm const} =
  {\textstyle\half}\Theta_6|_{\rm const} =
  -\frac{\zeta(2)\zeta(3)}{16}.
\end{equation}

Having determined all the coefficients, we can now write
\begin{align}
  J_{12|13} &= \frac{\Theta_4|_{\rm const}}{4\pi^2}v \notag\\
  &\quad + \frac{\Theta_5|_{\rm const}}{16\pi^2}(s^2+2w^2+su+3sw+tw+uv+2uw+2vw) \notag\\
  &\quad - \frac{\Theta_6|_{\rm const}}{32\pi^2}v^2
    + \frac{\Theta_5|_{\rm const}-\Theta_7|_{\rm const}}{16\pi^2}sv
    - \frac{\Theta_8|_{\rm const}}{16\pi^2}tv \notag\\
  &\quad + \mathcal{O}(\alpha'^3),
\end{align}
which now needs to be rewritten in terms of $k_r\cdot k_s$ and
generalised to $J_{rs|rt}$. For some terms this is trivial. For
example, it is clear that the $v$ multiplying $\Theta_4|_{\rm const}$
should be generalised to $-2k_s\cdot k_t$. Other terms are more tricky
although, by considering various other values for $r,\,s,\,t$, it can
be shown that, up to order $\alpha'^2$,
\begin{align} \label{eq:FinalJexp}
  J_{rs|rt} &= -\frac{\Theta_4|_{\rm const}}{2\pi^2}k_s\cdot k_t \notag\\
  &\quad + \frac{\Theta_5|_{\rm const}-\half\Theta_6|_{\rm const}}{4\pi^2} (k_s \cdot k_t)^2
     + \frac{\Theta_5|_{\rm const}}{4\pi^2} (k_r \cdot k_u)(k_r \cdot k_v) \notag\\
  &\quad + \frac{\Theta_5|_{\rm const}-\Theta_7|_{\rm const}}{4\pi^2}(k_r\cdot k_s)(k_s \cdot k_t)
    + \frac{\Theta_5|_{\rm const}-\Theta_8|_{\rm const}}{4\pi^2}(k_r\cdot k_t)(k_s \cdot k_t) \notag\\
  &\quad - \frac{\Theta_5|_{\rm const}}{4\pi^2}\big( (k_s \cdot k_u)(k_s \cdot k_v) +
  (k_s \cdot k_u)(k_t \cdot k_u) + (k_s \cdot k_u)(k_t \cdot k_v) \notag\\
  &\qquad\qquad\qquad + (k_s \cdot k_v)(k_t \cdot k_u) + (k_s \cdot k_v)(k_t \cdot k_v)
    +(k_t \cdot k_u)(k_t \cdot k_v) \big),
\end{align}
where $k_u$, $k_v$ are the two momenta other than $k_r$, $k_s$,
$k_t$. Only $\Theta_6$, $\Theta_7$ and $\Theta_8$ have non-zero
constant parts and so, using \eqref{eq:Theta78} and reinstating
$\alpha'$, this simplifies to
\begin{equation}
  J_{rs|rt} = \frac{\zeta(3)\alpha'^2}{2^5\cdot 3} (k_s\cdot
  k_t)(k_s\cdot k_t + k_r\cdot k_s + k_r\cdot k_t) + \mathcal{O}(\alpha'^3).
\end{equation}

\subsection{Expansion of Integral $J'$}
Integral $J'$ differs from $J$ in that all of the indices on the
$\eta\bar\eta$ are different. Specialising to the case $r,s=1,2$ and
$t,u=3,4$, we have
\begin{equation}
  J'_{12|34} = \frac{1}{\pi^2} \dT \dVdash \chiproddash
  \frac{\partial}{\partial v}(\ln\chi_{12})
  \frac{\partial}{\partial \bar v} (\ln\chi_{34}).
\end{equation}
Again, studying the region $v_{rs}\to 0$ shows the absence of massless
poles. This completes our earlier claim that the only source of
massless poles is integral $I$.

As with $J$, we begin with the most general possible expansion
given in \eqref{eq:Jexp2}. The details are identical in spirit to
those in section \ref{sec:Jexp} and are not presented here. It is
found that, up to and including order $\alpha'^2$, all coefficients
vanish except
\begin{equation}
  d_7 = -\frac{1}{16\pi^2}\Theta_5|_{\rm const},
  \qquad d_8 = \frac{1}{16\pi^2}\Theta_5|_{\rm const}.
\end{equation}
This is easily extended to general $r,\,s,\,t,\,u$, giving
\begin{equation}
  J'_{rs|tu} = -\frac{\Theta_5|_{\rm const}}{4\pi^2}\big( (k_r\cdot k_t)(k_s\cdot k_u)
  - (k_r\cdot k_u)(k_s\cdot k_t) \big).
\end{equation}
Since $\Theta_5$ has no non-zero constant part, we conclude that
\begin{equation}
  J'_{rs|tu} = 0 + \mathcal{O}(\alpha'^3).
\end{equation}
Although $J'$ vanishes at all orders considered in this paper, it is
not identically zero and will start to contribute at some higher
order. The effect of its vanishing up to order $\alpha'^2$ is that the
effective action can be written as $t_8t_8(D^{2n}R^3)(R^2)$, where the
final two Riemann tensors contract straight into the $t_8$ tensors. At
higher orders, when $J'$ starts to contribute, this will no longer be
the case.

\subsection{Expansion of Integral $I$}
As it stands, the low-energy expansion of $I$ makes little
sense. However, after removing the poles the integral become finite
for vanishing momenta and so can be expanded for small $k_r\cdot
k_s$. Consider the case where $r,s=1,2$,
\begin{equation}
  I_{12} = \frac{1}{\pi^2} \dT \dVdash \chiproddash
  \frac{\partial}{\partial{v}}(\ln\chi_{12})
  \frac{\partial}{\partial\bar{v}}(\ln\chi_{12}).
\end{equation}
Since the poles are known from section \ref{sec:Ipoles}, they can, in
principle, be subtracted order by order in $\alpha'$. Although some
progress can be made this way, it becomes increasingly difficult to
find suitable representations for the pole terms. Instead we use an
alternative method which allows $I$ to be directly expressed in terms
of the integrals $J$ and $K$.

Consider the following integral,
\begin{equation} \label{eq:Iint}
  \frac{1}{\pi^2} \dT \dVdash \, \frac{\partial}{\partial\bar v_1} \left(
  \chiproddash \frac{\partial}{\partial v}\ln\chi_{12} \right),
\end{equation}
which vanishes since $v_r$ is integrated over a surface with no boundary. (The
potential singularity near $|v_{12}|=0$ is easily shown to be finite
in the limit $s\to 0$.) By acting with the $\bar{v}_1$ derivative,
various $I$, $J$ and $K$ integrals are generated, leading to the
relation
\begin{equation}
  I_{12} = -\frac{k_1\cdot k_3}{k_1\cdot k_2}J_{12|13} -\frac{k_1\cdot
  k_4}{k_1\cdot k_2}J_{12|14} -\frac{k_1\cdot k_5}{k_1\cdot
  k_2}J_{12|15} - \frac{2}{k_1\cdot k_2}K,
\end{equation}
which clearly only contains single poles in $k_1\cdot k_2$. For
general $r$, $s$ the result is generalised to
\begin{equation}
  I_{rs} = -\sum_{t=1,\, t\neq r,s}^5 \frac{k_r\cdot k_t}{k_r\cdot
  k_s}J_{rs|rt} - \frac{1}{\alpha'k_r\cdot k_s}K,
\end{equation}
where $\alpha'$ has been reinserted. The expansion of $I$ is then
easily obtained using the expansions of $J$ and $K$ given in
\eqref{eq:FinalJexp} and \eqref{eq:FinalKexp} respectively. At order
$\alpha'^{-1}$ we find $\frac{1}{6\alpha'k_r\cdot k_s}$, agreeing
with the lowest order expansion of \eqref{eq:Ipoles2}. Orders
$\alpha'^0$ and $\alpha'^1$ vanish. The next order contribution, after
separating poles and non-poles, is
\begin{align}
  &\frac{\alpha'^2}{6\pi^2(k_r\cdot k_s)}\Bigg(
  \sum_{t\neq r,s}\Big(3(k_r\cdot k_t)(k_s\cdot k_t)^2\Theta_6|_{\rm const}
  +6(k_r\cdot k_t)^2(k_s\cdot k_t)\Theta_8|_{\rm const}\Big) \notag\\
  &\qquad\qquad\qquad +\sum_{(t,u)\neq(r,s)}(k_t\cdot k_u)^3\Theta_6|_{\rm const} \Bigg) \notag\\
  &\qquad\, +\frac{\alpha'^2}{6\pi^2}\Bigg(6\sum_{t\neq r,s}(k_r\cdot
  k_t)(k_s\cdot k_t)\Theta_7|_{\rm const}
  + (k_r\cdot k_s)^2\Theta_6|_{\rm const}\Bigg),
\end{align}
where we have not assumed any relationship between $\Theta_6$,
$\Theta_7$ and $\Theta_8$. From unitarity we know that the kinematic
factor multiplying the pole must involve the expansion of the
four-graviton amplitude to third order, which is
\begin{equation}
  \sum_{t\neq r,s}\big((k_r+k_s)\cdot k_t\big)^3 
  + \sum_{(t,u)\notin\{r,s\}}\big(k_t\cdot k_u\big)^3,
\end{equation}
where each sum contains three terms. Matching the two requires
$2\Theta_8|_{\rm const}=\Theta_6|_{\rm const}$, confirming the claim
in \eqref{eq:Theta78}. The remaining relationship of
\eqref{eq:Theta78}, which involves $\Theta_7$, is shown in the same
way but with $\partial_v$ interchanged with $\bar\partial_v$ in
\eqref{eq:Iint}.

Using \eqref{eq:Theta56const}, our final expression for the expansion
of $I_{rs}$ up to order $\alpha'^2$ is
\begin{align}
  I_{rs} &= \frac{1}{6\alpha'k_r\cdot k_s} \notag\\
  &\quad - \frac{\zeta(3)\alpha'^2}{2^5\cdot 3^2(k_r\cdot k_s)} \Bigg(
  \sum_{t\neq r,s}\big((k_r+k_s)\cdot k_t\big)^3 +
  \sum_{t,u\notin\{r,s\}}\big(k_t\cdot k_u\big)^3 \Bigg) \notag\\
  &\quad - \frac{\zeta(3)\alpha'^2}{2^5\cdot 3^2}\Bigg( 3\sum_{t\neq
  r,s}(k_r\cdot k_t)(k_s\cdot k_t) + (k_r\cdot k_s)^2 \Bigg),
\end{align}
where the first line is the lowest-order pole, the second line is the
order $\alpha'^2$ pole, and the third line is the order $\alpha'^2$
non-pole.

%%%%%%%%%%%%%%%%%%%%%%%%%%%%%%%%%%%%%%%%%%%%%%%%%%%%%%%%%%%%%%%%%%%%%%%%%%%%%%%%%%%%%%%%%%%%%%%%%%%%%%%%%%%%%%
\section{Consequences for the Effective Action} \label{sec:5gravEff}
%%%%%%%%%%%%%%%%%%%%%%%%%%%%%%%%%%%%%%%%%%%%%%%%%%%%%%%%%%%%%%%%%%%%%%%%%%%%%%%%%%%%%%%%%%%%%%%%%%%%%%%%%%%%%%
Given the low-energy expansion of the five-graviton amplitude, we can
now determine whether this implies new terms in the type II effective
action. As reviewed in section \ref{sec:4gravEffAct}, the one-loop
four-graviton amplitude implies the following one-loop terms in the
effective action up to order $D^6R^4$,
\begin{equation} \label{eq:KnownEA}
  \alpha'^3\int d^{10}x \sqrt{-g}\left(
  4\zeta(2)e^{\phi/2}\Rf + 0\cdot e^{-\phi/2}\alpha'^2D^4\Rf +
  8\zeta(2)\zeta(3)e^{-\phi}\alpha'^3D^6\Rf \right),
\end{equation}
where we are now using Einstein frame since this simplifies the
subsequent analysis\footnote{Unlike the string frame, the Einstein
frame contains no mixing between the graviton and dilaton
propagators.}. Using the four-graviton amplitude it is not possible to
determine exactly how the derivatives are distributed amongst the
Riemann tensors; for concreteness we assume $D^{2n}\Rf$ is shorthand
for
\begin{equation}
  t_8^{a_1b_1a_2b_2a_3b_3a_4b_4} t_8^{c_1d_1c_2d_2c_3d_3c_4d_4}
  ((D^2)^n(R_{a_1b_1c_1d_1}R_{a_2b_2c_2d_2}))\,
  R_{a_3b_3c_3d_3}R_{a_4b_4c_4d_4}.  
\end{equation}
By studying the five-graviton amplitude we can address three
issues. Firstly, we can resolve the question about how the derivatives
are distributed (modulo the $D^2R$ issue discussed below). Secondly,
we can determine whether a $D^2R^4$ term exists, which cannot be seen
from the four-graviton amplitude. Thirdly, we can study whether it is
necessary to add new $R^5$, $D^2R^5$ and $D^4R^5$ terms.

The strategy will be to calculate the contribution to the
five-graviton amplitude from \eqref{eq:KnownEA} by expanding around
flat space and considering all possible tree-level diagrams. After
removing these diagrams, any remaining terms will be covariantised to
find potentially novel $D^{2n}R^5$ terms.

\subsection{Ambiguities in the Effective Action}
As mentioned in the introduction, on-shell effective actions can only
be determined up to field redefinitions. Since the Weyl tensor differs
from the Riemann tensor by terms involving the Ricci tensor and Ricci
scalar, it is impossible to distinguish the two, and $R^4$ can be
replaced by $W^4$. This conclusion only holds if other fields, such as
the \NSNS\ two-form and the \RR\ fields, are turned off. If they are
not, then it is still true that $R^4$ and $W^4$ can be interchanged,
but only at the expense of adding additional terms involving the other
fields. Consequently, it may well be the case that either $R^4$ or
$W^4$ is preferred since it leads to an effective action with fewer
terms. For the case here, with all other fields turned off, we choose
to use the Riemann tensor since its expansion around flat space is
considerably simpler.

Now consider terms involving $D^2R_{abcd}$. By using the Bianchi
identity and replacing $D_{[a}D_{b]}$ by Riemann tensors, it can be
shown that
\begin{align} \label{eq:D2RisRR}
  D^2R_{abcd} &= 2{R^e}_{afb}{R^f}_{ecd} - 2{R^e}_{afc}{R^f}_{deb} +
  2{R^e}_{afd}{R^f}_{ceb} \notag\\
  &\quad + R_{ae}{R^e}_{bcd} - R_{be}{R^e}_{acd} \notag\\
  &\quad + D_aD_cR_{db} - D_aD_dR_{cb} - D_bD_cR_{da} + D_bD_dR_{ca},
\end{align}
and so, after removing the Ricci terms using a field redefinition,
$D^2R_{abcd}$ can be replaced by a sum of Riemann-squared terms. This
implies, for example, that $D^2\Rf$ can be replaced by a sum of $R^5$
terms (which is one explanation for why $D^2\Rf$ does not contribute
to the four-graviton amplitude). Similarly
\begin{equation}
  (D_fD_eR)(D^eD^fR^3) = (D_eD_fR)(D^eD^fR^3) + \mbox{``$D^4R^5$
  terms''},
\end{equation}
and so the issue of how the derivatives are distributed in
$D^{2n}\Rf$ is actually ill-defined: different distributions are often
equivalent up to $R^5$ terms. Then the only possible criteria for
fixing the precise meaning of $D^{2n}\Rf$ involves choosing the term
which leads to the fewest total number of terms in the effective
action.

\subsection{Expansions of Various Tensors}
Before expanding \eqref{eq:KnownEA}, we first need to expand the
Riemann, Ricci and $t_8$ tensors around flat space. Consider a small
fluctuation of the metric about the Minkowski metric,
\begin{equation}
  g_{ab} = \eta_{ab} + \kappa h_{ab},
\end{equation}
where $\kappa$ is presumed small. In subsequent expressions we will
drop factors of $\kappa$ since they can easily be reinstated; the
order of the expansion is then given by the number of $h$
factors. Indices on $h$ and $\partial$ are raised and
lowered with $\eta$, whereas all other indices are raised and lowered
with $g$. So, for example,
\begin{equation}
  {\Gamma^a}_{bc} = g^{ad}\Gamma_{dbc}, \qquad {h^a}_b =
  \eta^{ac}h_{cb}, \qquad \partial^a = \eta^{ab}\partial_b.
\end{equation}
The expansions of the inverse metric and the metric determinant are
readily found to be
\begin{align}
   g^{ab} &= \eta^{ab} - h^{ab} + h^{ac}{h_c}^b + \cdots, \label{eq:invgexp} \\
  \sqrt{-g} &= 1 + {\textstyle\half} {h^a}_a + \cdots, \label{eq:gExp}
\end{align}
and likewise for the Christoffel symbol,
\begin{equation}
  {\Gamma^a}_{bc} = {\textstyle\half}(\partial_b {h^a}_c
  + \partial_c {h^a}_b - \partial^a h_{bc})
   - {\textstyle\half}h^{ad}(\partial_b h_{cd} + \partial_c h_{bd} -
   \partial_d h_{bc}) + \cdots.
\end{equation}

Since the expansion of $D^{2n}\Rf$ is required up to fifth order in
$h$, we need the expansion of $R_{abcd}$ up to second order. After a
slightly more involved calculation, it can be shown that
\begin{align} \label{eq:RiemannExp}
  \eta^{ce}\eta^{df}R_{abef} &= 2\partial_{[a} \partial^{[c} {h_{b]}}^{d]} \notag\\
  &\quad + \partial_{[a} h^{e[c} \partial^{d]} h_{b]e} +
    {\textstyle\half}\partial_{[a} h^{ec} \partial_{b]} {h^d}_e +
    {\textstyle\half}\partial^{[c} h_{ae} \partial^{d]} {h_b}^e \notag\\
  &\quad - \partial^e {h_{[a}}^{[c} \partial_{b]} {h^{d]}}_e -
    \partial^e {h_{[a}}^{[c} \partial^{d]} h_{b]e} +
    {\textstyle\half}\partial^e {h_a}^{[c} \partial_e {h^{d]}}_b.
\end{align}
Often the Riemann tensor $R_{abcd}$ is multiplied by
$t_8^{ab\cdots}t_8^{cd\cdots}$ which is antisymmetric in
$a\leftrightarrow b$ and in $c\leftrightarrow d$. This allows the
first term to be rewritten as $2\partial_a \partial_c h_{bd}$, without
the square brackets. Further, there is also a symmetry in
$(a,\,b)\leftrightarrow(c,\,d)$. This symmetry is not 
obvious, but, since the $\Rf$ term is only expanded to fifth order, it
follows that at least three Riemann tensors are only expanded to first
order which, even when written as $2\partial_a \partial_c h_{bd}$,
still has manifest $(a,\,b)\leftrightarrow(c,\,d)$ symmetry. With the
understanding that $R_{abcd}$ is multiplied by a tensor with these
symmetries, its expansion to second order simplifies to
\begin{equation} \label{eq:RiemannExp2}
  R_{abcd} = 2\partial_a \partial_c h_{bd}
    + \partial_a h_c^{\phantom{c}e} \partial_d h_{be} +
    \partial_a h_c^{\phantom{c}e} \partial_b h_{de}
     - 2\partial^e h_{ac} \partial_b h_{de}
     + {\textstyle\half} \partial^e h_{ac} \partial_e h_{bd}.
\end{equation}

By contracting \eqref{eq:RiemannExp} with an inverse metric, the
expansion of the Ricci tensor can be determined as
\begin{align}
  R_{ab} &= {\textstyle\half}(\Box h_{ab} + \partial_a\partial_bh
    - \partial_a\partial_c{h^c}_b - \partial_b\partial_c{h^c}_a) \notag\\
  &\quad - {\textstyle\half} h^{cd}(\partial_a\partial_bh_{cd} + \partial_c\partial_dh_{ab}
    - \partial_a\partial_ch_{bd} - \partial_b\partial_ch_{ad}) \notag\\
  &\quad - {\textstyle\quart}\partial_ah_{cd}\partial_bh^{cd}
    + {\textstyle\half}(\partial^ch_{ad}\partial^dh_{bc}-\partial_ch_{ad}\partial^c{h^d}_b) \notag\\
  &\quad + {\textstyle\quart}(\partial_ah_{bc}+\partial_bh_{ac}-\partial_ch_{ab})
    (2\partial_dh^{cd}-\partial^ch),
\end{align}
where $h={h^a}_a$ and $\Box=\partial^a\partial_a$. Similarly, the
Ricci scalar is given by
\begin{align} \label{eq:RExp}
  R &= \Box h - \partial_a\partial_b h^{ab} \notag\\
  &\quad - h^{ab}(\Box h_{ab} + \partial_a\partial_bh
    - 2\partial_a\partial^ch_{bc}) \notag\\
  &\quad - {\textstyle\frac{3}{4}}\partial_ah_{bc}\partial^ah^{bc}
    + {\textstyle\half}\partial_ah_{bc}\partial^bh^{ac}
    + \partial^ah_{ab}\partial_ch^{bc} - \partial^ah_{ab}\partial^bh
    + {\textstyle\quart}\partial^ah\partial_ah.
\end{align}
We will actually need the Ricci scalar expanded to third order in
$h$. This has previously been calculated in
\cite{DeWitt:1967uc} and there is no need to reproduce the result here.

Finally, it is important to remember that the $t_8$ tensor must also
be expanded. For the amplitudes calculated here, amplitudes in flat space,
$t_8$ appears as a sum of products of inverse Minkowski
metrics. However, when written in an effective action, $t_8$ is
covariantised to involve full metrics and, as such, should be expanded
in $h$. Where it is important to distinguish $t_8$ written in terms of
the Minkowski metric from $t_8$ written in terms of the full metric we
define
\begin{align} \label{eq:underlinet8}
  t_8^{abcdefgh} &= \sum g^{\cdot\cdot}g^{\cdot\cdot}g^{\cdot\cdot}g^{\cdot\cdot} =
  -\half g^{ac}g^{bd}g^{eg}g^{fh} + \mbox{59 other terms}, \notag\\
  \underline{t}_8^{abcdefgh} &= \sum \eta^{\cdot\cdot}\eta^{\cdot\cdot}\eta^{\cdot\cdot}\eta^{\cdot\cdot} =
  -\half\eta^{ac}\eta^{bd}\eta^{eg}\eta^{fh} + \mbox{59 other terms}.
\end{align}
Then, to first order in $h$, $t_8$ can be expanded in terms of
$\underline{t}_8$ as
\begin{align}
  t_8^{abcdefgh} &= \underline{t}_8^{abcdefgh} \notag\\
  &\quad - {\textstyle\half}( {h_i}^a\underline{t}_8^{ibcdefgh} +
    {h_i}^b\underline{t}_8^{aicdefgh} +
    {h_i}^c\underline{t}_8^{abidefgh} +
    {h_i}^d\underline{t}_8^{abciefgh} \notag\\
  &\quad\qquad + h_i^{\phantom{i}e}\underline{t}_8^{abcdifgh} +
    {h_i}^f\underline{t}_8^{abcdeigh} +
    {h_i}^g\underline{t}_8^{abcdefih} +
    {h_i}^h\underline{t}_8^{abcdefgi} ).
\end{align}
As with the Riemann tensor, $t_8$ is often multiplied by a tensor
which is symmetric under the interchange of pairs of indices, \eg
under $(a,b) \leftrightarrow (c,d)$. For example, when $\Rf$ is
expanded to fifth order, the lowest order expansion of the Riemann
tensor, as shown in \eqref{eq:RiemannExp2}, has this symmetry. Then
the expansion of $t_8$ simplifies to
\begin{equation} \label{eq:t8exp}
  t_8^{abcdefgh} = \underline{t}_8^{abcdefgh} -2 (
  {h_i}^a\underline{t}_8^{ibcdefgh} +
  {h_i}^b\underline{t}_8^{aicdefgh} ).
\end{equation}

\subsection{Expansion of the Known Effective Action}
We are now in a position to expand \eqref{eq:KnownEA} up to
fifth order. This will give the graviton propagator and various
three-, four- and five-point vertices, some of which will be
associated with tree-level terms and some with one-loop terms.

\subsubsection{The Einstein-Hilbert Term}
Consider first the Einstein-Hilbert action, $\int d^{10}x\sqrt{-g}
R$. The first contribution, using \eqref{eq:gExp} and \eqref{eq:RExp}
and dropping total derivatives, is at second order,
\begin{equation}
  S_{EH} = \quart\int d^{10}x\, (\partial_ah_{bc}\partial^ah^{bc} -
  \partial_ah\partial^ah +2\partial_ah\partial_bh^{ab} -2
  \partial_ah_{bc}\partial^bh^{ac}),
\end{equation}
which is invariant under the gauge transformation
\begin{equation}
  h_{ab} \to h_{ab} + \partial_a\zeta_b + \partial_b\zeta_a,
\end{equation}
where $\zeta_a$ is an arbitrary one-form field. We fix the gauge
invariance using the de Donder gauge, $\partial^ah_{ab} =
{\textstyle\half}\partial_bh$, which leads to the usual graviton
propagator,
\begin{equation}
  D_{ab,cd} = \frac{\eta_{ac}\eta_{bd} + \eta_{ad}\eta_{bc} -
  \quart\eta_{ab}\eta_{cd}}{k^2}.
\end{equation}

The expansion of $\sqrt{-g}R$ to third order gives a
three-graviton vertex, $V_{3\,ab,cd,ef}^R$, which was first calculated in
\cite{DeWitt:1967uc}. The result is not given here since we will take
a short-cut when calculating diagrams involving such a vertex.

\subsubsection{The $\Rf$ Term} \label{sec:R4exp}
Since each Riemann tensor contains at least one $h$, the expansion of
$\int d^{10}x\sqrt{-g}\,\Rf$ begins at fourth order. With all tensors
expanded to lowest order, the action becomes
\begin{equation} \label{eq:R4exp}
  S^{\Rf}_{4h} = 2^4 \int d^{10}x\,\underline{t}_8^{a_1b_1\cdots}
    \underline{t}_8^{c_1d_1\cdots} \partial_{a_1}\partial_{c_1}h_{b_1d_1}\,
    \partial_{a_2}\partial_{c_2}h_{b_2d_2}\, \partial_{a_3}\partial_{c_3}h_{b_3d_3}\,
    \partial_{a_4}\partial_{c_4}h_{b_4d_4}
\end{equation}
and it is straightforward to read off the four-graviton vertex as
\begin{align} \label{eq:R44vertex}
  V^{\Rf}_{4\,ab,cd,ef,gh} &= 2^4\,\underline{t}_8^{a_1b_1a_2b_2a_3b_3a_4b_4}
  \underline{t}_8^{c_1d_1c_2d_2c_3d_3c_4d_4}\, k_{1\,a_1}k_{1\,c_1}
  k_{2\,a_2}k_{2\,c_2} k_{3\,a_3}k_{3\,c_3} k_{4\,a_4}k_{4\,c_4} \notag\\
  &\qquad\times \eta_{b_1a}\eta_{d_1b} \eta_{b_2c}\eta_{d_2d}
  \eta_{b_3e}\eta_{d_3f} \eta_{b_4g}\eta_{d_4h}.
\end{align}

Expanding to fifth order is more involved. The fifth graviton can
originate either from the $\sqrt{-g}$, from a $t_8$ tensor or from a Riemann
tensor. When it originates from the $\sqrt{-g}$, the term will
necessarily involve a $\half h$ factor, which can be ignored for our
purposes since this vertex will only ever be used in diagrams with all
legs on-shell. With this understanding,
\begin{align} \label{eq:R45vertex}
  S^{\Rf}_{5h} &= 2^5 \int d^{10}x\,\underline{t}_8^{a_1b_1a_2b_2a_3b_3a_4b_4}
    \underline{t}_8^{c_1d_1c_2d_2c_3d_3c_4d_4}
    \partial_{a_2}\partial_{c_2}h_{b_2d_2}\,\partial_{a_3}\partial_{c_3}h_{b_3d_3}\,
    \partial_{a_4}\partial_{c_4}h_{b_4d_4} \notag\\
  &\qquad\times ( -\partial_{a_1}{h_{d_1}}^e\partial_{c_1}h_{b_1e} +
    \partial_{a_1}{h_{c_1}}^e\partial_{b_1}h_{d_1e} -
    2\partial^eh_{b_1d_1}\partial_{a_1}h_{c_1e} \notag\\
  &\qquad\qquad\qquad + {\textstyle\half}\partial^eh_{a_1c_1}\partial_eh_{b_1d_1} -
    2h_{c_1e}\partial_{a_1}\partial^eh_{b_1d_1} - 
    2h_{b_1e}\partial_{c_1}\partial_{a_1}{h_{d_1}}^e ),
\end{align}
where the first four terms originate from expanding a Riemann tensor
to second order, and the final two terms are from expanding a $t_8$
tensor. The relevant five-vertex is easily read off.

\subsubsection{The $D^6\Rf$ Term} \label{sec:D6R4exp}
There is no need to expand $D^2\Rf$ and $D^4\Rf$ since the amplitude
vanishes at these orders. However, there is a non-vanishing
contribution at order $D^6R^4$. To be explicit, we assume $D^6\Rf$ is
shorthand for
\begin{equation}
  t_8^{a_1b_1\cdots}t_8^{c_1d_1\cdots}
  D_eD_fD_gR_{a_1b_1c_1d_1} \, D^eD^fD^gR_{a_2b_2c_2d_2} \,
  R_{a_3b_3c_3d_3} \, R_{a_4b_4c_4d_4}.
\end{equation}
Since at lowest order the covariant derivatives become ordinary
derivatives, the first contribution to the expansion is very similar
to that for $\Rf$,
\begin{align}
  S^{D^6\Rf}_{4h} &= 2^4 \int d^{10}x\,
    \underline{t}_8^{a_1b_1a_2b_2a_3b_3a_4b_4}
    \underline{t}_8^{c_1d_1c_2d_2c_3d_3c_4d_4}\notag\\
  &\qquad\times \partial_e\partial_f\partial_g
    \partial_{a_1}\partial_{c_1}h_{b_1d_1}\,
    \partial^e\partial^f\partial^g\partial_{a_2}\partial_{c_2}h_{b_2d_2}\,
    \partial_{a_3}\partial_{c_3}h_{b_3d_3}\,
    \partial_{a_4}\partial_{c_4}h_{b_4d_4}.
\end{align}
However, to fifth order there is the added complication of expanding the
covariant derivatives. The fifth graviton can now either come from the
$\sqrt{-g}$, a $t_8$ tensor, a Riemann tensor, a Christoffel symbol
within a covariant derivative, or an inverse metric used to raise an
index on the second set of derivatives. After a certain amount of work
and a little rearranging, it can be shown that
\begin{align} \label{eq:D6R4to5}
  S^{D^6\Rf}_{5h} &= 2^4 \int d^{10}x\,\underline{t}_8^{a_1b_1a_2b_2a_3b_3a_4b_4}
    \underline{t}_8^{c_1d_1c_2d_2c_3d_3c_4d_4} \notag\\
  &\quad\times \raisebox{0pt}[15pt][0pt]{\bigg(}
    \partial_e\partial_f\partial_g( -\partial_{a_1}{h_{d_1}}^k\partial_{c_1}h_{b_1k} +
    \partial_{a_1}{h_{c_1}}^k\partial_{b_1}h_{d_1k} -
    2\partial^kh_{b_1d_1}\partial_{a_1}h_{c_1k} \notag\\
  &\quad\qquad\qquad\qquad + {\textstyle\half}\partial^kh_{a_1c_1}\partial_kh_{b_1d_1} -
    2\partial_{a_1}\partial^kh_{b_1d_1} h_{c_1k} - 
    2\partial_{c_1}\partial_{a_1}{h_{d_1}}^kh_{b_1k} ) \notag\\
  &\quad\qquad\qquad\times \partial^e\partial^f\partial^g
    \partial_{a_2}\partial_{c_2}h_{b_2d_2}\, \partial_{a_3}\partial_{c_3}h_{b_3d_3}\,
    \partial_{a_4}\partial_{c_4}h_{b_4d_4} \notag\\
  &\qquad\quad + ( -\partial_{a_1}{h_{d_1}}^k\partial_{c_1}h_{b_1k} +
    \partial_{a_1}{h_{c_1}}^k\partial_{b_1}h_{d_1k} -
    2\partial^kh_{b_1d_1}\partial_{a_1}h_{c_1k} \notag\\
    &\quad\qquad\qquad\qquad + {\textstyle\half}\partial^kh_{a_1c_1}\partial_kh_{b_1d_1} -
    2\partial_{a_1}\partial^kh_{b_1d_1} h_{c_1k} - 
    2\partial_{c_1}\partial_{a_1}{h_{d_1}}^kh_{b_1k} ) \notag\\
  &\quad\qquad\qquad\times \partial_e\partial_f\partial_g
    \partial_{a_2}\partial_{c_2}h_{b_2d_2}\,
    \partial^e\partial^f\partial^g\partial_{a_3}\partial_{c_3}h_{b_3d_3}\,
    \partial_{a_4}\partial_{c_4}h_{b_4d_4} \notag\\
  &\qquad\quad - \Big(2\partial_e\partial_f(\partial_{a_1}{h_g}^k-\partial^kh_{a_1g})
    (\partial_k\partial_{c_1}h_{b_1d_1}-\partial_{b_1}\partial_{c_1}h_{kd_1}) \notag\\
  &\qquad\qquad\quad + 6\partial_e(\partial_{a_1}{h_g}^k-\partial^kh_{a_1g})
    \partial_f(\partial_k\partial_{c_1}h_{b_1d_1}-\partial_{b_1}\partial_{c_1}h_{kd_1}) \notag\\
  &\qquad\qquad\quad + 6(\partial_{a_1}{h_g}^k-\partial^kh_{a_1g})
    \partial_e\partial_f(\partial_k\partial_{c_1}h_{b_1d_1}-\partial_{b_1}\partial_{c_1}h_{kd_1}) \notag\\
  &\qquad\qquad\quad + (2\partial_e\partial_f{h_g}^k-\partial_e\partial^kh_{fg})
    \partial_k\partial_{a_1}\partial_{c_1}h_{b_1d_1} \notag\\
  &\qquad\qquad\quad + 3(2\partial_e{h_g}^k-\partial^kh_{eg})
    \partial_f\partial_k\partial_{a_1}\partial_{c_1}h_{b_1d_1} \notag\\
  &\qquad\qquad\quad - 3{h_e}^k\partial_f\partial_g\partial_k
    \partial_{a_1}\partial_{c_1}h_{b_1d_1} \Big) \notag\\
  &\quad\qquad\qquad\times \partial^e\partial^f\partial^g
    \partial_{a_2}\partial_{c_2}h_{b_2d_2}\, \partial_{a_3}\partial_{c_3}h_{b_3d_3}\,
    \partial_{a_4}\partial_{c_4}h_{b_4d_4}
    \raisebox{0pt}[0pt]{\bigg)},
\end{align}
where in the top half the fifth graviton originates from a Riemann tensor
or a $t_8$ tensor, and in the bottom half from a covariant derivative
or an inverse metric. Again, terms involving $h$ have been ignored.

\subsection{Diagrams from the $\Rf$ Term} \label{sec:R4diagrams}
\begin{figure}
\begin{center}
\includegraphics[scale=0.8]{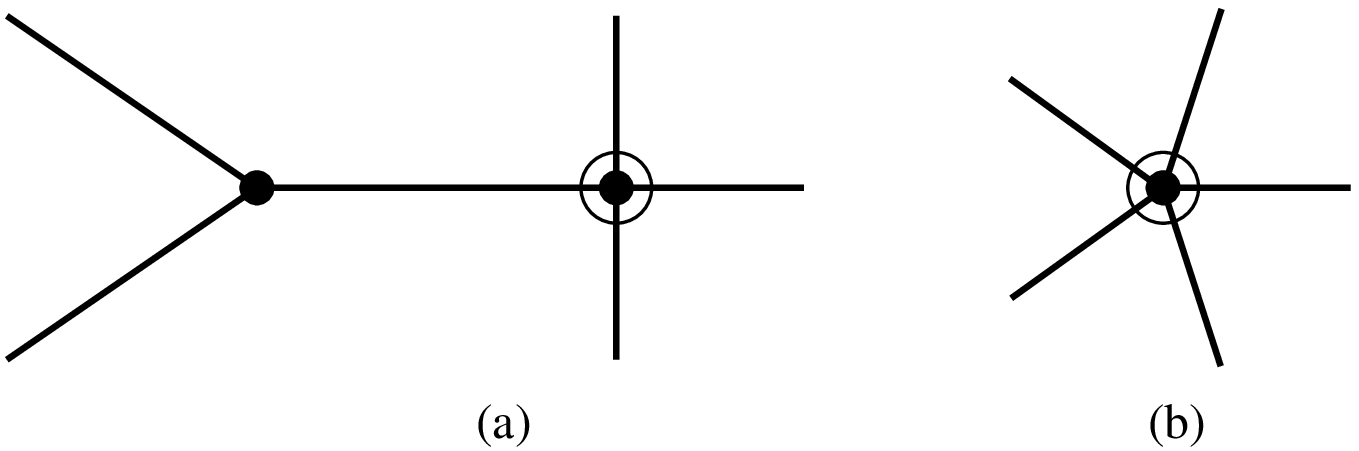}
\caption{The two field theory diagrams contributing to the
five-graviton amplitude: (a) a pole diagram and (b) a contact
diagram.}
\label{fig:EffActDiags}
\end{center}
\end{figure}

Now we calculate the relevant diagrams which contribute to the
five-graviton amplitude. Since diagrams involving $D^6\Rf$ are similar
in spirit to those involving $\Rf$, we focus only on the latter. As
explained in the introduction, only tree-level diagrams need be
considered and, since we are studying a one-loop amplitude, exactly
one of the vertices must originate from a one-loop term. Such terms
begin at $\Rf$ and so each diagram must contain either a one-loop
four-vertex or a one-loop five-vertex. This leads to only two diagrams
as shown in figure \ref{fig:EffActDiags}, where a dot represents a
vertex from the expansion of the Einstein-Hilbert action, and a circle
surrounding a dot represents a vertex from an $\Rf$ term.

Diagram (a) contains a three-vertex from the expansion of $R$
connected via a graviton propagator to a four-vertex from the
expansion of $\Rf$. This diagram will be responsible for the poles in
the amplitude, although it will also contain non-pole pieces where the
pole in the denominator is cancelled by the numerator. Diagram (b) is
simply the $\Rf$ five-vertex contracted into five on-shell external
gravitons.

In calculating these diagrams, we will focus only on the `$s$-channel'
since all other cases work in an identical manner. For diagram (a) the
meaning of this is clear: the incoming particles on the left
are particles $1$ and $2$ carrying momenta $k_1$ and $k_2$
respectively. However, perhaps counter-intuitively, diagram (b) can
also be split into different channels as follows. All terms in the
$\Rf$ five-vertex \eqref{eq:R45vertex} contain a
$\partial_{a_2}\partial_{c_2}h_{b_2d_2}
\partial_{a_3}\partial_{c_3}h_{b_3d_3}
\partial_{a_4}\partial_{c_4}h_{b_4d_4}$
factor multiplied by two other gravitons. The two other gravitons
originate either from expanding a Riemann tensor to second order or by
expanding a $t_8$ tensor. By `$s$-channel' we mean choosing these two
other gravitons to be particles $1$ and $2$.

First consider figure \ref{fig:EffActDiags}(a). To evaluate this
diagram we take the Einstein-Hilbert three-vertex and contract into
two external particles, numbers $1$ and $2$, and take the $\Rf$
four-vertex and contract into three external particles, numbers $3$,
$4$ and $5$. Then we sandwich the two together using a graviton
propagator. This whole procedure is quite involved, largely due to the
complicated nature of the three-vertex. However, we can take a
short-cut since an almost identical calculation was performed in
\cite{Sannan:1986tz}, which considers four-graviton scattering
at tree-level and  matches with the low-energy limit of the
same amplitude in string theory. Similar in spirit to figure
\ref{fig:EffActDiags}, \cite{Sannan:1986tz} contains two diagrams: a
pole diagram and a contact diagram. The pole diagram involves an
Einstein-Hilbert three-vertex connected via a graviton propagator to
another Einstein-Hilbert three-vertex. As such, the first part of the
calculation is identical to the one considered here, the only
difference being that here the second vertex is a four-vertex from the
$\Rf$ term.

Taking the three-vertex and contracting two legs into on-shell
gravitons gives (3.2) in \cite{Sannan:1986tz} which, after further
contracting into a propagator, results in (3.7). This result is for the
$t$-channel, but is easily converted to the $s$-channel
giving
\begin{align}
  &h_1^{ab}h_2^{cd}V^R_{3\,ab,cd,ef}(k_1,k_2,-k_1-k_2)D^{ef,mn}(k_1+k_2) = \notag\\
  &\qquad\qquad (h_1h_2)^{mn} - \frac{1}{s}\Big( (h_1h_2)k_1^mk_1^n
    + 2(h_1h_2)k_1^{(m}k_2^{n)} + (h_1h_2)k_2^mk_2^n \notag\\
  &\qquad\qquad + (k_1h_2k_1)h_1^{mn} + (k_2h_1k_2)h_2^{mn}
    - 4(k_1h_2h_1)_{\phantom{1}}^{(m}k_1^{n)} \notag\\
  &\qquad\qquad - 4(k_2h_1h_2)_{\phantom{2}}^{(m}k_2^{n)}
    - 4(k_1h_2)^{(m}(k_2h_1)^{n)} \Big),
\end{align}
where $(h_1h_2)=h_{1,ab}h_2^{ab}$, $(h_1h_2)_{ab}=h_{1,ac}{h_{2,b}}^c$
and so on. At this point we deviate from the calculation in
\cite{Sannan:1986tz} and instead contract into the four-vertex given
in \eqref{eq:R44vertex}, with the remaining three legs contracted into
gravitons $3$, $4$ and $5$. After using the antisymmetry of $t_8$ we
obtain
\begin{align} \label{eq:R4PoleDiagram}
  \parbox{9mm}
  {\includegraphics[scale=0.3]{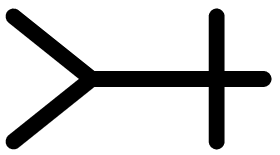}}
  &= h^1_{a_1c_1}h^2_{a_2c_2}h^3_{a_3c_3}h^4_{a_4c_4}h^5_{a_5c_5}\,
     k_3^{b_3}k_4^{b_4}k_5^{b_5} k_3^{d_3}k_4^{d_4}k_5^{d_5} \notag\\
  & \qquad\times\raisebox{0pt}[19pt][0pt]{\Bigg(} -\frac{1}{s} \Big(
     k_1^{a_2}(k_1+k_2)^bt_8^{a_1ba_3b_3a_4b_4a_5b_5}
    -k_2^{a_1}(k_1+k_2)^bt_8^{a_2ba_3b_3a_4b_4a_5b_5} \notag\\
  & \qquad\qquad\qquad\quad
    -\delta^{a_1a_2}k_1^{b_1}k_2^{b_2}t_8^{b_1b_2a_3b_3a_4b_4a_5b_5}
    \Big) \notag\\
  & \qquad\qquad \times\Big(
    k_1^{c_2}(k_1+k_2)^dt_8^{c_1dc_3d_3c_4d_4c_5d_5}
    -k_2^{c_1}(k_1+k_2)^dt_8^{c_2dc_3d_3c_4d_4c_5d_5} \notag\\
  & \qquad\qquad\qquad\quad 
    -\delta^{c_1c_2}k_1^{d_1}k_2^{d_2}t_8^{d_1d_2c_3d_3c_4d_4c_5d_5}
    \Big) \notag\\
  & \qquad\qquad+\delta^{c_1c_2}(k_1+k_2)^b(k_1+k_2)^d
    t_8^{a_1ba_3b_3a_4b_4a_5b_5} t_8^{a_2dc_3d_3c_4d_4c_5d_5}
    \raisebox{0pt}[0pt]{\Bigg)},
\end{align}
where all terms are poles except the final line.

Evaluating diagram \ref{fig:EffActDiags}(b) is simpler. We take the
five-vertex from \eqref{eq:R45vertex} and contract into the five
gravitons, considering all relevant permutations of the external
particles. For the `$s$-channel' this means permuting gravitons $1$
and $2$ between the $\partial h\partial h$ and $h\partial\partial h$
terms and permuting gravitons $3$, $4$ and $5$ between the remaining
$\partial\partial h$ terms, which gives
\begin{align} \label{eq:R4ContactDiagram}
  \parbox{6mm}
  {\includegraphics[scale=0.3]{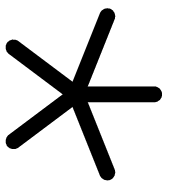}}
  &= h^1_{a_1c_1}h^2_{a_2c_2}h^3_{a_3c_3}h^4_{a_4c_4}h^5_{a_5c_5}\,
     k_3^{b_3}k_4^{b_4}k_5^{b_5} k_3^{d_3}k_4^{d_4}k_5^{d_5} \notag\\
  & \qquad\times\Big(\delta^{c_1c_2}k_1^{b_1}k_2^{b_2}t_8^{b_1b_2a_3b_3a_4b_4a_5b_5}
    t_8^{a_1a_2c_3d_3c_4d_4c_5d_5} \notag\\
  & \qquad\qquad -\delta^{c_1c_2}k_1^{b_1}k_2^{d_2}t_8^{a_2b_1a_3b_3a_4b_4a_5b_5}
    t_8^{a_1d_2c_3d_3c_4d_4c_5d_5} \notag\\
  & \qquad\qquad -k_1^{c_2}k_2^{b_2}t_8^{c_1b_2a_3b_3a_4b_4a_5b_5}
    t_8^{a_1a_2c_3d_3c_4d_4c_5d_5} \notag\\
  & \qquad\qquad\qquad +k_1^{b_1}k_2^{c_1}t_8^{c_2b_1a_3b_3a_4b_4a_5b_5}
    t_8^{a_1a_2c_3d_3c_4d_4c_5d_5} \notag\\
  & \qquad\qquad -{\textstyle\frac{1}{4}}s\,t_8^{a_1a_2a_3b_3a_4b_4a_5b_5}
    t_8^{c_1c_2c_3d_3c_4d_4c_5d_5} \notag\\
  & \qquad\qquad -k_1^{c_2}k_1^{b_1}t_8^{c_1b_1a_3b_3a_4b_4a_5b_5}
    t_8^{a_1a_2c_3d_3c_4d_4c_5d_5} \notag\\
  & \qquad\qquad\qquad +k_2^{c_1}k_2^{b_2}t_8^{c_2b_2a_3b_3a_4b_4a_5b_5}
    t_8^{a_1a_2c_3d_3c_4d_4c_5d_5} \notag\\
  & \qquad\qquad -\delta^{c_1c_2}k_1^{b_1}k_1^{d_1}t_8^{a_1b_1a_3b_3a_4b_4a_5b_5}
    t_8^{a_2d_1c_3d_3c_4d_4c_5d_5} \notag\\
  & \qquad\qquad\qquad -\delta^{c_1c_2}k_2^{b_2}k_2^{d_2}t_8^{a_1b_2a_3b_3a_4b_4a_5b_5}
    t_8^{a_2d_2c_3d_3c_4d_4c_5d_5} \Big),
\end{align}
where indented lines show the $1\leftrightarrow 2$ permutations.

\subsection{Matching with the Effective Action}
These diagrams, and the equivalent diagrams at higher orders, can now
be subtracted from the expansion of the amplitude in
section \ref{sec:5gravExp}. Any remaining terms can then be
covariantised to discover possible $R^5$ and $D^{2n}R^5$ terms.

\subsubsection{Order $R^4$} \label{sec:R4Matching}
At lowest order the amplitude contains eight powers of momenta and so
is relevant to the $\Rf$ term. Since the $I$ and $J$ integrals are
multiplied by ten powers of momenta, whereas $K$ is only
multiplied by eight, we need to consider the expansions of $I$
and $J$ at order $\alpha^{-1}$ and the expansion of $K$ at order
$\alpha^0$,
\begin{equation}
  I_{rs}|_{\alpha'^{-1}} = \frac{1}{6\alpha'k_r\cdot k_s}, \qquad
  J_{rs|rt}|_{\alpha'^{-1}} = 0, \qquad J'_{rs|tu}|_{\alpha'^{-1}} = 0,
  \qquad K|_{\alpha'^0} = -\frac{1}{6}.
\end{equation}
Since we are just considering the `$s$-channel', the amplitude
\eqref{eq:amp} to lowest order is given solely by the
$\eta_{12}\bar\eta_{12}$ and $\hat\Omega_{12}$ terms,
\begin{align}
  A_{5h,t_8t_8}|_{\Rf} &= \frac{2^5\alpha'^4}{3} h^1_{a_1c_1}h^2_{a_2c_2}h^3_{a_3c_3}h^4_{a_4c_4}h^5_{a_5c_5}\,
    k_3^{b_3}k_4^{b_4}k_5^{b_5} k_3^{d_3}k_4^{d_4}k_5^{d_5} \notag\\
  & \qquad\times\Bigg( -\frac{1}{s} \Big( k_1^{a_2}(k_1+k_2)^{b}t_8^{a_1ba_3b_3a_4b_4a_5b_5}
    -k_2^{a_1}(k_1+k_2)^{b}t_8^{a_2ba_3b_3a_4b_4a_5b_5} \notag\\
  & \qquad\qquad\qquad\quad -\delta^{a_1a_2}k_1^{b_1}k_2^{b_2}t_8^{b_1b_2a_3b_3a_4b_4a_5b_5}
    +{\textstyle\half}s t_8^{a_1a_2a_3b_3a_4b_4a_5b_5}\Big) \notag\\
  & \qquad\qquad\quad\times \Big( k_1^{c_2}(k_1+k_2)^dt_8^{c_1dc_3d_3c_4d_4c_5d_5}
    -k_2^{c_1}(k_1+k_2)^dt_8^{c_2dc_3d_3c_4d_4c_5d_5} \notag\\
  & \qquad\qquad\qquad\quad -\delta^{c_1c_2}k_1^{d_1}k_2^{d_2}t_8^{d_1d_2c_3d_3c_4d_4c_5d_5}
    +{\textstyle\half}s t_8^{c_1c_2c_3d_3c_4d_4c_5d_5}\Big) \notag\\
  & \qquad\qquad +\delta^{a_1c_2}\, k_1^{d_1}k_2^{b_2}
  t_8^{a_2b_2a_3b_3a_4b_4a_5b_5}t_8^{c_1d_1c_3d_3c_4d_4c_5d_5} \Bigg),
\end{align}
where $\alpha'$ has been reinstated using $2\alpha'=1$. This is
identical to the sum of the two field theory diagrams,
\eqref{eq:R4PoleDiagram} and \eqref{eq:R4ContactDiagram}. Of course,
this has to be the case since $R^5$ terms only begin to contribute at
the next order; the usual $\Rf$ term has to account for the full
amplitude at this order.

\subsubsection{Order $D^2R^4$}
At the next order we are looking for possible $D^2R^4$ terms, which
start to contribute at the same order as $R^5$ terms. The relevant
terms in the expansions of the modular integrals all vanish,
\begin{equation}
  I_{rs}|_{\alpha'^0} = 0, \qquad J_{rs|rt}|_{\alpha'^0} = 0, \qquad
  J'_{rs|tu}|_{\alpha'^0} = 0, \qquad K|_{\alpha'^1} = 0,
\end{equation}
which implies the vanishing of all $D^2\Rf$ and $R^5$ terms.

It is worth noting that any $D^2\Rf$ term can always be rewritten as a
sum of $R^5$ terms as follows. There are two possible $D^2\Rf$ terms
depending on whether or not the covariant derivatives act on the same
Riemann tensor:
\begin{align}
  &(D^2R_{a_1b_1c_1d_1})R_{a_2b_2c_2d_2}
  R_{a_3b_3c_3d_3}R_{a_4b_4c_4d_4}, \notag\\
  &(D^eR_{a_1b_1c_1d_1})(D_eR_{a_2b_2c_2d_2})
  R_{a_3b_3c_3d_3}R_{a_4b_4c_4d_4}.
\end{align}
However, since these differ by a total derivative, we are free to
consider just the first which, from \eqref{eq:D2RisRR}, is
equivalent to a sum of $R^5$ terms. The converse is clearly not true:
most $R^5$ terms cannot be re-expressed as $D^2\Rf$ terms. So we have
shown that all $R^5$ terms, including the combination equivalent to
$D^2\Rf$, vanish. Such a conclusion cannot be drawn from studying the
four-graviton amplitude: the fact that $D^2\Rf$ is equivalent to $R^5$
terms shows that it does not contribute at four gravitons. The
five-graviton amplitude establishes that $D^2\Rf$ really is absent.

This result agrees with the expectation in the literature (for example
in \cite{Frolov:2001xr}). Also it complements the results in
\cite{Grisaru:1986wj} where it was shown that there is no
contribution to the {\it tree-level} gravitational $\beta$-function at
five loops and that any $R^5$ term must vanish on a K\"ahler
manifold.

This analysis only applies to terms involving the $t_8$ tensor. Terms
with $\epsilon_{10}$ tensors will be studied in the next section.

\subsubsection{Order $D^4R^4$}
The next order contains terms such as $D^4R^4$, $D^2R^5$ and $R^6$,
although the presence of the latter can only be determined by studying
amplitudes with at least six gravitons. Again, the expansions of the
modular integrals vanish,
\begin{equation} \label{eq:D4R4exp}
  I_{rs}|_{\alpha'^1} = 0, \qquad J_{rs|rt}|_{\alpha'^1} = 0, \qquad
  J'_{rs|tu}|_{\alpha'^1} = 0, \qquad K|_{\alpha'^2} = 0,
\end{equation}
implying the absence of $D^4\Rf$ terms. This agrees with the
conclusion from the four-graviton amplitude where, as shown
in section \ref{sec:4gravB}, the $s^2+t^2+u^2$ coefficient in the
low-energy expansion vanishes.

Unlike the previous order, it is no longer true that all $D^4\Rf$
terms can be rewritten as $D^2R^5$ terms. If, for simplicity, we
ignore the order of the indices on the derivatives, then of the seven
arrangements of the four derivatives over the Riemann tensors, only
three are independent; the others are equal up to total
derivatives. These can be taken to be
\begin{equation}
  (D_eD_fR)(D^eD^fR)R^2, \qquad (D_eD^eD_fD^fR)R^3, \qquad
  (D_eD^eR)(D_fD^fR)R^2.
\end{equation}
The first is a true $D^4\Rf$ term in that it cannot be rewritten as a
sum of $D^2R^5$ terms and would contribute to the four-graviton
amplitude. The second can, using \eqref{eq:D2RisRR}, be written as a
sum of $D^2R^5$ terms which would contribute to five gravitons but not
to four. Using \eqref{eq:D2RisRR} twice, the third is equivalent to a
sum of $R^6$ terms and therefore makes no contribution to either the
four- or five-graviton amplitudes. From \eqref{eq:D4R4exp}, we
conclude that the first two terms are both absent, although we can say
nothing about the third.

Not only have we shown that $(D_eD^eD_fD^fR)R^3\sim\sum D^2R^5$ vanishes,
but that most other $D^2R^5$ terms are also zero, where by most we mean
all terms which cannot be rewritten as $R^6$ terms. An example of a
term which cannot be determined is $(D_eR)(D^eR)R^3$. This follows
from the same reason that $(D_eR)(D^eR)R^2$ cannot be determined using
the four-graviton amplitude. If we use the convention that terms are
always written using the greatest possible number of Riemann tensors,
then we have shown that all $D^4R^4$ and $D^2R^5$ terms vanish.

\subsubsection{Order $D^6R^4$} \label{sec:D6R4Matching}
Finally we consider terms at order $D^6R^4\sim D^4R^5$, which is as
far as can be studied using the expansions in section
\ref{sec:5gravExp}. It will be possible to determine both the most
appropriate arrangement of derivatives in the $D^6\Rf$ term and to
find new $D^4R^5$ terms. The relevant terms in the expansions of $I$,
$J$, $J'$ and $K$ are given by
\begin{align} \label{eq:D6R4exp}
  I_{rs}|_{\alpha'^2} &=
    - \frac{\zeta(3)\alpha'^2}{2^5\cdot 3^2(k_r\cdot k_s)} \Bigg(
    \sum_{t\neq r,s}\big((k_r+k_s)\cdot k_t\big)^3 +
    \sum_{t,u\notin\{r,s\}}\big(k_t\cdot k_u\big)^3
    \Bigg) \notag\\
  &\qquad\quad - \frac{\zeta(3)\alpha'^2}{2^5\cdot 3^2} \Bigg(
    3\sum_{t\neq r,s}(k_r\cdot k_t)(k_s\cdot k_t) + (k_r\cdot k_s)^2
    \Bigg), \notag\\
  J_{rs|rt}|_{\alpha'^2} &= \frac{\zeta(3)\alpha'^2}{2^5\cdot 3} (k_s\cdot
    k_t)(k_s\cdot k_t + k_r\cdot k_s + k_r\cdot k_t), \notag\\
  J'_{rs|tu}|_{\alpha'^2} &= 0, \notag\\
  K|_{\alpha'^3} &= \frac{\zeta(3)\alpha'^3}{2^5\cdot 3^2} \sum_{r<s}
    (k_r\cdot k_s)^3.
\end{align}
Since $J_{rs|rt}|_{\alpha'^2}$ is non-zero, we can no longer just
consider the `$s$-channel'. Instead we use the following prescription
to reduce the number of diagrams which must be calculated. For terms
where two particles are singled out, these are chosen to be particles
$1$ and $2$; for terms where three particles are singled out, these
are chosen as particles $1$, $2$ and $3$, with particle $3$ occupying
the repeated position. So, to be explicit, we only need to consider
the $|\eta_{12}|^2$, $\eta_{13}\bar\eta_{23}$,
$\eta_{23}\bar\eta_{13}$ and $\hat\Omega_{12}$ terms in the
amplitude.

The $D^6\Rf$ term in \eqref{eq:KnownEA} leads to two diagrams, which
are identical in spirit to those considered in section
\ref{sec:R4diagrams}, the only difference being that the one-loop
vertices now originate from $D^6\Rf$ rather than from $\Rf$. The
calculation of the pole diagram proceeds exactly as in section
\ref{sec:R4diagrams}, the final result for the $s$-channel being
\eqref{eq:R4PoleDiagram} multiplied by
\begin{align} \label{eq:D6ksum}
  &\big((k_1+k_2)\cdot k_3\big)^3 + \big((k_1+k_2)\cdot k_4\big)^3
  + \big((k_1+k_2)\cdot k_5\big)^3 \notag\\
  &\quad + \big(k_3\cdot k_4\big)^3 + \big(k_3\cdot k_5\big)^3
  + \big(k_4\cdot k_5\big)^3.
\end{align}
The contact diagram is evaluated by contracting \eqref{eq:D6R4to5}
into five on-shell gravitons, with particles $1$ and $2$ for the first
two gravitons, particle $3$ for the third, and particles $4$
and $5$ for the final two. There is no need to write the
result here since it is essentially \eqref{eq:D6R4to5} with
derivatives replaced by momenta.

Both these diagrams need to be subtracted from the amplitude before
the remaining terms can be covariantised. It is helpful to first
remove terms containing \eqref{eq:D6ksum} since these exactly mirror
the matching at order $\Rf$: the pole diagram and the first half of
the contact diagram \eqref{eq:D6R4to5} are easily seen to match with
the $\frac{1}{k_1\cdot k_2}$ terms from $I_{12}$ and the equivalent
terms from $K$ in \eqref{eq:D6R4exp}. Of course, the pole terms have
to match since, by unitarity, the can be understood as arising from a
four-graviton amplitude connected by an on-shell graviton to a
three-graviton amplitude. This leaves the second half of
\eqref{eq:D6R4to5} to be subtracted from the remaining terms in
\eqref{eq:D6R4exp}. After tedious but straightforward work all the
extra terms in \eqref{eq:D6R4to5} are found in the amplitude with the
exception of
\begin{align} \label{eq:extraD6term}
  &-2^5 \int d^{10}x\,\underline{t}_8^{a_1b_1a_2b_2a_3b_3a_4b_4}
    \underline{t}_8^{c_1d_1c_2d_2c_3d_3c_4d_4}\,
    \partial_e\partial_f(\partial_{a_1}{h_g}^k-\partial^kh_{a_1g}) \notag\\
  &\qquad\quad\times(\partial_k\partial_{c_1}h_{b_1d_1}-\partial_{b_1}\partial_{c_1}h_{kd_1})
    \, \partial^e\partial^f\partial^g
    \partial_{a_2}\partial_{c_2}h_{b_2d_2}\, \partial_{a_3}\partial_{c_3}h_{b_3d_3}\,
    \partial_{a_4}\partial_{c_4}h_{b_4d_4}.
\end{align}
However, by considering the expansion of
\begin{align}
  &\bar{t}_{10}^{\,ABCDa_2b_2a_3b_3a_4b_4}t_8^{c_1d_1c_2d_2c_3d_3c_4d_4}\,
    \partial_e\partial_f\partial_Ah_{gB} \,
    \partial^e\partial^f\partial^g\partial_C\partial_{c_1}h_{Dd_1} \notag\\
  &\qquad\times \partial_{a_2}\partial_{c_2}h_{b_2d_2}
    \, \partial_{a_3}\partial_{c_3}h_{b_3d_3}
    \, \partial_{a_4}\partial_{c_4}h_{b_4d_4}
\end{align}
in terms of $t_8$ tensors and using \eqref{eq:bart10identity}, it can
be shown that \eqref{eq:extraD6term} is equivalent to
\begin{align}
  &\frac{2^5}{3} \int d^{10}x\,\underline{t}_8^{a_1b_1a_2b_2a_3b_3a_4b_4}
    \underline{t}_8^{c_1d_1c_2d_2c_3d_3c_4d_4}\,
    \partial_e\partial_f(\partial_{a_1}{h_g}^k-\partial^kh_{a_1g}) \notag\\
  &\qquad\times \partial^e\partial^f\partial^g
    (\partial_k\partial_{c_1}h_{b_1d_1}-\partial_{b_1}\partial_{c_1}h_{kd_1})
    \, \partial_{a_2}\partial_{c_2}h_{b_2d_2} \, \partial_{a_3}\partial_{c_3}h_{b_3d_3} \,
    \partial_{a_4}\partial_{c_4}h_{b_4d_4},
\end{align}
which does match with terms in the $(k_1\cdot k_2)^2$ part of
$I_{12}$.

The remaining terms in the amplitude require new $D^4R^5$ terms in the
effective action to reproduce them. It is a simple matter to
covariantise these extra terms finding
\begin{align} \label{eq:D4R5terms}
  S_{D^4R^5} &= 8\zeta(3)\zeta(2) \alpha'^6 \int
  d^{10}x\sqrt{-g}\,e^{-\phi}\,t_8^{a_1b_1a_2b_2a_3b_3a_4b_4}
  t_8^{c_1d_1c_2d_2c_3d_3c_4d_4} \notag\\
  &\qquad\times \Big( 12 D_eR_{a_1fc_1g}D_k
  R^{f\phantom{b_1}g}_{\phantom{f}b_1\phantom{g}d_1}
  D^eD^kR_{a_2b_2c_2d_2} R_{a_3b_3c_3d_3}
  R_{a_4b_4c_4d_4} \notag\\
  &\qquad\qquad +4 D_eD_fR_{a_1gc_1k}D^eD^f
  R^{g\phantom{b_1}k}_{\phantom{g}b_1\phantom{k}d_1}
  R_{a_2b_2c_2d_2} R_{a_3b_3c_3d_3} R_{a_4b_4c_4d_4} \notag\\
  &\qquad\qquad +12 D^2 (D^eR_{a_1fc_1d_1} D_eR_{a_2b_2gd_2}
  R^{f\phantom{b_1c_2}g}_{\phantom{f}b_1c_2}) R_{a_3b_3c_3d_3}
  R_{a_4b_4c_4d_4} \notag\\
  &\qquad\qquad +6 D^2
  (D^eR_{a_1fc_1d_1}R_{a_2b_2ge}D^gR^f_{\phantom{f}b_1c_2d_2})
  R_{a_3b_3c_3d_3} R_{a_4b_4c_4d_4} \notag\\
  &\qquad\qquad -24 R_{a_1efg} D^fD_kR^e_{\phantom{e}b_1c_1d_1}
  D^gD^kR_{a_2b_2c_2d_2} R_{a_3b_3c_3d_3} R_{a_4b_4c_4d_4} \notag\\
  &\qquad\qquad -3 D^2 (D_eD_fR_{a_1b_1c_1d_1}
  R^{\phantom{a_2b_2}f}_{a_2b_2\phantom{f}g} R^{eg}_{\phantom{eg}c_2d_2})
  R_{a_3b_3c_3d_3} R_{a_4b_4c_4d_4} \notag\\
  &\qquad\qquad +3 D_eR_{a_1b_1fg} D_kR^{fg}_{\phantom{fg}c_1d_1}
  D^eD^kR_{a_2b_2c_2d_2} R_{a_3b_3c_3d_3}
  R_{a_4b_4c_4d_4} \notag\\
  &\qquad\qquad +{\textstyle\frac{1}{3}} D_eD_fR_{a_1b_1gk}
  D^eD^fR^{gk}_{\phantom{gk}c_1d_1} R_{a_2b_2c_2d_2} R_{a_3b_3c_3d_3}
  R_{a_4b_4c_4d_4} \Big),
\end{align}
where the normalisation has been chosen to agree with the convention
in section \ref{sec:4gravEffAct}. This is to be added to the $D^6\Rf$
term known from studying the four-graviton amplitude,
\begin{align} \label{eq:D6R4term}
  S_{D^6\Rf} &= 8\zeta(3)\zeta(2) \alpha'^6 \int
  d^{10}x\sqrt{-g}\,e^{-\phi}\,t_8^{a_1b_1a_2b_2a_3b_3a_4b_4}
  t_8^{c_1d_1c_2d_2c_3d_3c_4d_4} \notag\\
  &\qquad\times  D_eD_fD_gR_{a_1b_1c_1d_1}
  D^eD^fD^g R_{a_2b_2c_2d_2}
  R_{a_3b_3c_3d_3} R_{a_4b_4c_4d_4}.
\end{align}

We can now address the issue of the `most appropriate' arrangement of
derivatives in $D^6\Rf$. Because, using \eqref{eq:D2RisRR},
we can exchange $D^2R$ terms for $R^2$ terms, the `correct' $D^6\Rf$
term is an ill-defined concept and so `most appropriate' can only
refer to the number of terms. We seek the particular $D^6\Rf$ term
which requires the fewest number of extra $D^4R^5$ terms. For example,
perhaps $D^2(D_eD_fR_{a_1b_1c_1d_1} D^eD^fR_{a_2b_2c_2d_2})
R_{a_3b_3c_3d_3} R_{a_4b_4c_4d_4}$ requires fewer extra terms than
\eqref{eq:D6R4term}. In other words, we ask whether some of the
$D^4R^5$ terms in \eqref{eq:D4R5terms} can be rewritten as $D^6\Rf$
terms. However, there is no obvious way this can be done. No group of
terms has the correct form to be rewritten either as $D^2R_{abcd}$ or
as $D_aD_bR_{efcd}$, and permuting indices on derivatives (\ie
considering $D_fD_eR$ rather than $D_eD_fR$) leads to no obvious
improvement. So we conclude that \eqref{eq:D6R4term}, the combination
hinted at from the four-graviton amplitude, is the `most appropriate'
$D^6\Rf$ term, with the extra $D^4R^5$ terms given by
\eqref{eq:D4R5terms}.

\subsection{Various Conjectures}
It is notable that the coefficient in front of the one-loop $D^6\Rf$ term
arising from a four-graviton calculation is identical to the
coefficient in front of the one-loop $D^4R^5$ term arising from a
five-graviton calculation. This is despite arising, at least
superficially, from the expansion of totally different modular
integrals. The same is also true at the two previous orders, where the
coefficients for both the four- and five-graviton amplitudes vanish.
It is quite plausible that this behaviour persists at tree-level,
which leads to various conjectures for the five-graviton {\it
tree-level} amplitude. In particular, using the expansion of the
four-graviton tree-level amplitude \cite{Green:1999pv}, we conjecture
that the low-energy expansion of the equivalent five-graviton
amplitude will be $\frac{1}{16}\zeta(5)$ at order $\alpha'^5$ and
$\frac{1}{96}\zeta(3)^2$ at order $\alpha'^6$ multiplied, in each
case, by the same kinematic factor as at one-loop.

Further, as reviewed in section \ref{sec:IIBEffAct}, for IIB string
theory it is possible to extend the tree- and one-loop four-graviton
results to all orders in the string coupling, even non-perturbatively,
finding modular functions such as the non-holomorphic Eisenstein
series $Z_{3/2}(\tau,\bar\tau)$. Since the one-loop terms in these
series also match with the five-graviton amplitude (at least up to
order $D^6\Rf$), a bold conjecture is that the same modular function
that multiplies $D^{2n}\Rf$ also multiplies the corresponding
$D^{2n-2}R^5$ term. This then allows various higher-loop five-graviton
amplitude conjectures in IIB. Firstly, at order $\alpha'^3$, we
conjecture no loop corrections above one-loop. Secondly, at order
$\alpha'^5$, we expect a two-loop contribution but nothing higher. In
particular, motivated by the two-loop coefficient in
\eqref{eq:IIBD4R4}, we conjecture that the five-graviton two-loop
amplitude contains a $\frac{4}{45}\pi^2$ factor at this
order. Finally, at order $\alpha'^6$, the two- and three-loop
five-graviton amplitudes are conjectured to be non-zero, with the
coefficients matching the four-graviton coefficients in
\eqref{eq:IIBD6R4}, and with all higher-order perturbative amplitudes
vanishing.

For IIB one can even go a step further and conjecture that the modular
function in front of $D^{2n}R^4$ is not only the same as the function
in front of $D^{2n-2}R^5$, but is in fact quite universal and
multiplies all terms at the same order. For example, at order
$\alpha'^6$, perhaps the $\mathcal{E}_{(3/2,3/2)}(\tau,\bar\tau)$
which multiplies the $D^6\Rf$ term also multiplies the $D^4R^5$,
$D^2R^6$ and $R^7$ terms.

%%%%%%%%%%%%%%%%%%%%%%%%%%%%%%%%%%%%%%%%%%%%%%%%%%%%%%%%%%%%%%%%%%%%%%%%%%%%%%%%%%%%%%%%%%%%%%%%%%%%%%%%%%%%%%
\section{The $\epsilon_8\epsilon_8$ Terms} \label{sec:5gravEps}
%%%%%%%%%%%%%%%%%%%%%%%%%%%%%%%%%%%%%%%%%%%%%%%%%%%%%%%%%%%%%%%%%%%%%%%%%%%%%%%%%%%%%%%%%%%%%%%%%%%%%%%%%%%%%%
So far we have implicitly ignored terms in the amplitude involving
$\epsilon_8$ tensors. These appear in the trace over four $R_0^{ab}$
factors, which can be written as the sum of an $\epsilon_8$ and sixty
$\delta\delta\delta\delta$ terms as in \eqref{eq:t8tensor},
\begin{align} \label{eq:trR4}
   \tr (R^{ab}_0 R^{cd}_0 R^{ef}_0 R^{gh}_0)
   &= \pm\half\epsilon_8^{abcdefgh}
   -\half\delta^{ac}\delta^{bd}\delta^{eg}\delta^{fh}
   + \cdots \notag\\
   &\equiv \pm\half\epsilon_8^{abcdefgh} + t_8^{abcdefgh},
\end{align}
with the $\pm$ sign depending on the $SO(8)$ chirality of the $S$
fields. There is often ambiguity in the literature as to whether $t_8$
is defined as the whole expression or just as the sum of delta
symbols, and we have so far avoided this issue. From now on we choose
the second definition, as in \eqref{eq:trR4}. All effective actions
written in previous sections should be interpreted in this way. The
precise definition makes little difference for the four-graviton
amplitude since the $\epsilon_8$ terms vanish due to momentum
conservation. However, this is no longer true for the five-graviton
case.

The calculation of the full five-graviton amplitude including
$\epsilon_8$ tensors proceeds exactly as in section \ref{sec:amp}. Any
occurrence of a $t_8$ tensor originates from a trace over four
$R_0^{ab}$ tensors and so can more generally be replaced by
\eqref{eq:trR4}. The only non-trivial part involves checking that
\eqref{eq:bart10identity} continues to hold when $t_8$ is replaced by
$\epsilon_8$, which can be shown using \eqref{eq:8DGram}. So, the full
amplitude is given by \eqref{eq:amp} with $t_8$ everywhere replaced by
$\pm\half\epsilon_8+t_8$. The arguments demonstrating modular
invariance, gauge invariance and Bose symmetry proceed as before.

Although the form of \eqref{eq:amp} is unchanged, certain terms
simplify when $t_8$ is replaced by $\epsilon_8$. In particular, the
first three lines of \eqref{eq:A12} all vanish. This is easily seen if
$k_1$ is replaced by $-(k_2+k_3+k_4+k_5)$, after which all but the
final line contain vanishing $k_r^ak_r^b\epsilon_8^{\cdots a\cdots
b\cdots}$ factors. However, it will be economical to ignore these
cancellations since then the results for $t_8$ tensors can be directly
applied to $\epsilon_8$ tensors.

After replacing the $t_8t_8$ factor in \eqref{eq:amp} by
$(t_8+\half\epsilon_8)(t_8\pm\half\epsilon_8)$, where the $\pm$ sign
will be discussed below, three different tensor structures appear:
$t_8t_8$, $\epsilon_8t_8$ and $\epsilon_8\epsilon_8$. The
$\epsilon_8t_8$ terms must vanish since, if they were present, they
would lead to terms in the effective action of the form
$\epsilon_{10}t_8D^nR^m$. Such terms, however, are odd under a
spacetime parity transformation and so cannot be present in either the
type IIA or type IIB theories. Checking this explicitly is quite
involved since it requires understanding the expansions of the modular
integrals to all orders in $\alpha'$. At lowest order, however, it can
be shown after a little work by using momentum conservation,
\eqref{eq:bart10identity} and \eqref{eq:8DGram}.

So, the amplitude reduces to $(t_8t_8\pm\quart\epsilon_8\epsilon_8)$
multiplied by the usual kinematic factors and integrals. The
$\epsilon_8\epsilon_8$ term is significantly simpler than the $t_8t_8$
part as a consequence of the cancellations mentioned above,
\begin{align} \label{eq:eeamp}
  A_{5h,\epsilon_8\epsilon_8} &= \pm\quart
  h^1_{a_1c_1}h^2_{a_2c_2}h^3_{a_3c_3}h^4_{a_4c_4}h^5_{a_5c_5}
  \dT \dV \chiprod \notag\\
  & \qquad\times \left( \sum_{r<s}\eta(v_{rs},\tau)A'_{rs}
  \sum_{r<s}\bar\eta(v_{rs},\tau)\bar{A}'_{rs}
  + \sum_{r<s}\hat\Omega(v_{rs},\tau)B'_{rs} \right),
\end{align}
where
\begin{equation}
  A'_{12} = k_1\cdot k_2\,k_3^{b_3}k_4^{b_4}k_5^{b_5}
  \,\epsilon_8^{a_1a_2a_3b_3a_4b_4a_5b_5}
\end{equation}
and $B'_{12}$ is as in \eqref{eq:B12} but with both $t_8$ tensors
replaced by $\epsilon_8$. Since $A'_{rs}$ always involves a $k_r\cdot
k_s$ factor, the potential pole from the $I_{rs}$ integral is always
cancelled, and so $A_{5h,\epsilon_8\epsilon_8}$ contains no massless
poles. This is directly related to the fact that, although, as
reviewed below, the effective action contains an
$\epsilon_{10}\epsilon_{10}R^4$ piece, this term does not lead to a
four-graviton vertex and so pole diagrams analogous to figure
\ref{fig:EffActDiags}(a) are guaranteed to vanish.

There is an issue with the sign of the $\epsilon_8\epsilon_8$ term
relative to the $t_8t_8$ term for both type IIA and IIB. For IIB, $S$
and $\tilde{S}$ have the same $SO(8)$ chirality and so the amplitude
contains a $(t_8\pm\half\epsilon_8) (t_8\pm\half\epsilon_8) \to
(t_8t_8+\quart\epsilon_8\epsilon_8)$ factor. Similarly, $S$ and
$\tilde{S}$ have opposite chiralities in IIA and so the tensor
structure is given by $(t_8\pm\half\epsilon_8)(t_8\mp\half\epsilon_8)
\to (t_8t_8-\quart\epsilon_8\epsilon_8)$. However, these cannot be the
correct signs for the $\epsilon_8\epsilon_8$ parts. As discussed
below, covariant calculations show that the flip of sign between IIA
and IIB is correct, but that the IIA theory should have the plus sign
and IIB the minus sign. The difference is presumably due either to
some limitation of the light-cone gauge GS formalism (the same issue
does not occur in the RNS approach) or to some unknown subtlety
relevant to the $\epsilon_8\epsilon_8$ terms. From now on we will
assume that this problem has been resolved and present the
$\epsilon_8\epsilon_8$ terms with the opposite signs, despite no
direct understanding of this from the current formalism.

Now that we have the $\epsilon_8\epsilon_8$ terms, we can ask whether
they lead to new $\epsilon_{10}\epsilon_{10}$ terms in the effective
action and, if so, how they package together with the $t_8t_8$
terms. First we review the known $t_8t_8$ and
$\epsilon_{10}\epsilon_{10}$ terms. Both the tree-level and one-loop
$t_8t_8R^4$ terms were discovered using the four-graviton amplitude
\cite{Gross:1986iv}; they have identical kinematic structure in both
IIA and IIB. However, such a calculation cannot reveal the presence of
$\epsilon_{10}\epsilon_{10}R^4$ terms where, to be precise,
\begin{equation}
  \epsilon_{10}\epsilon_{10}R^4\equiv
  \epsilon_{10\,mn}^{\phantom{10\,mn}a_1b_1\cdots
  a_4b_4}\epsilon_{10}^{mnc_1d_1\cdots c_4d_4}
  R_{a_1b_1c_1d_1}R_{a_2b_2c_2d_2}R_{a_3b_3c_3d_3}R_{a_4b_4c_4d_4}.
\end{equation}
At tree-level these terms where found by studying the four-loop beta
functions of the sigma model world-sheet action \cite{Grisaru:1986px,
Grisaru:1986vi, Freeman:1986zh, Park:1987jp}; again they were found to
have identical structures for IIA and IIB. For the equivalent one-loop
terms there are two famous calculations: \cite{Kiritsis:1997em}
considers string theory compactified on a two-torus and calculates a
five-point RNS amplitude involving four gravitons and a modulus field
of $T^2$; and \cite{Antoniadis:1997eg} compactifies on a
six-dimensional Calabi-Yau and calculates the three-graviton
amplitude, again in the RNS formalism. Both find the same structure as
at tree-level but with an important sign-flip for IIA. At one-loop in
IIA there is also the $\epsilon_{10}t_8BR^4$ term found in
\cite{Vafa:1995fj}, which we ignore here. So schematically,
suppressing the numerical coefficients, the IIB effective action in
Einstein frame is given by
\begin{equation} \label{eq:IIBeps}
  \alpha'^3\int d^{10}x\sqrt{-g}\,\Big(
  \underbrace{e^{-\frac{3}{2}\phi}
  (t_8t_8+{\textstyle\frac{1}{8}}\epsilon_{10}\epsilon_{10})R^4}_{\rm tree}
  +\underbrace{e^{\frac{1}{2}\phi}
  (t_8t_8+{\textstyle\frac{1}{8}}\epsilon_{10}\epsilon_{10})R^4}_{\rm 1-loop}
  \Big),
\end{equation}
whereas the equivalent for IIA is
\begin{equation} \label{eq:IIAeps}
  \alpha'^3\int d^{10}x\sqrt{-g}\,\Big(
  \underbrace{e^{-\frac{3}{2}\phi}
  (t_8t_8+{\textstyle\frac{1}{8}}\epsilon_{10}\epsilon_{10})R^4}_{\rm tree}
  +\underbrace{e^{\frac{1}{2}\phi}
  (t_8t_8-{\textstyle\frac{1}{8}}\epsilon_{10}\epsilon_{10})R^4}_{\rm 1-loop}
  \Big).
\end{equation}
Of course, it is important that there is no sign flip between
tree-level and one-loop in IIB since this would ruin the
$SL(2,\mathbb{Z})$ invariance. It will be possible, using the
$\epsilon_8\epsilon_8$ terms of the five-graviton amplitude, to
confirm both $\epsilon_{10}\epsilon_{10}R^4$ one-loop terms, and to
determine the presence of certain higher-order terms, such as
$\epsilon_{10}\epsilon_{10}D^6R^4$ and
$\epsilon_{10}\epsilon_{10}D^4R^5$.

To find these terms we need to expand the one-loop
$\epsilon_{10}\epsilon_{10}R^4$ term in small fluctuations about flat
space and calculate the same diagrams as in section
\ref{sec:R4diagrams}, but with $t_8t_8$ vertices replaced by
$\epsilon_{10}\epsilon_{10}$ vertices. The only new tensor which must
be expanded is $\epsilon_{10}\epsilon_{10}$. This is easily achieved
using
\begin{equation}
  {\epsilon_{10\,mn}^{}}^{a_1b_1\cdots a_4b_4}
  {{\epsilon_{10}}^{mn}}_{c_1d_1\cdots c_4d_4} = 2!8!(-1)^s\,\delta^{[a_1}_{c_1}
  \delta^{b_1}_{d_1}\cdots\delta^{a_4}_{c_4}\delta^{b_4]}_{d_4},
\end{equation}
where $s$, the number of minuses in the signature, is $1$ here.
This will contain inverse metrics after all the $c$ and $d$ indices
are raised. As with the $t_8$ tensor, we need to consider two different
$\epsilon_{10}\epsilon_{10}$ tensors: one formed out of the full
metric, $g$, and one formed out of the Minkowski metric,
$\eta$. Analogously to \eqref{eq:underlinet8}, we distinguish the two
by labelling the latter
$\underline{\epsilon}_{10}\underline{\epsilon}_{10}$. For comparison
with \eqref{eq:eeamp}, we require the contribution from
$\epsilon_{10}\epsilon_{10}$ in light-cone gauge. Since $k^+$ and
$h^{+i}$ are both zero in this gauge, the only non-zero contribution
arises from $m,n$ taking values $+,-$, leading to
\begin{align}
  \epsilon_{10}\epsilon_{10} &= 2!8!(-1)^s\,\sum g^{\cdot\cdot}\cdots
  g^{\cdot\cdot} \to -2\epsilon_8\epsilon_8, \notag\\
  \underline{\epsilon}_{10}\underline{\epsilon}_{10}
  &= 2!8!(-1)^s\,\sum\eta^{\cdot\cdot}\cdots\eta^{\cdot\cdot} \to
  -2\underline{\epsilon}_8\underline{\epsilon}_8,
\end{align}
where the minus signs arise since the $(+,-)$ light-cone coordinates
parametrize a space of Lorentzian signature\footnote{So, to be
explicit, with all indices contracted,
${\underline{\epsilon}_{10}}^{a\cdots}\underline{\epsilon}_{10\,a\cdots}=-10!$ and
$\underline{\epsilon}_8^{a\cdots}\underline{\epsilon}_{8\,a\cdots}=+8!$.}.
Using \eqref{eq:invgexp}, the expansion of
$\epsilon_{10}\epsilon_{10}$ to first order in $h$ can be shown to be
\begin{align}
  {\epsilon_{10\,mn}^{}}^{a_1b_1\cdots a_4b_4}
  {\epsilon_{10}}^{mnc_1d_1\cdots c_4d_4} &\to -2
  \underline{\epsilon}_8^{a_1b_1\cdots a_4b_4}
  \underline{\epsilon}_8^{c_1d_1\cdots c_4d_4} \notag\\
  &\qquad + (
  {h_i}^{a_1}\underline{\epsilon}_8^{ib_1\cdots a_4b_4}
    \underline{\epsilon}_8^{c_1d_1\cdots c_4d_4} +
  {h_i}^{b_1}\underline{\epsilon}_8^{a_1i\cdots a_4b_4}
    \underline{\epsilon}_8^{c_1d_1\cdots c_4d_4} \notag\\
  &\qquad\qquad + \cdots +
  {h_i}^{d_4}\underline{\epsilon}_8^{a_1b_1\cdots a_4b_4}
    \underline{\epsilon}_8^{c_1d_1\cdots c_4i} ).
\end{align}
If this is multiplied by the usual symmetries of $R^4$ it simplifies
to
\begin{align}
  {\epsilon_{10\,mn}^{}}^{a_1b_1\cdots a_4b_4}
  {\epsilon_{10}}^{mnc_1d_1\cdots c_4d_4} &\to -2
  \underline{\epsilon}_8^{a_1b_1\cdots a_4b_4}
  \underline{\epsilon}_8^{c_1d_1\cdots c_4d_4} \notag\\
  &\qquad + 8(
  {h_i}^{a_1}\underline{\epsilon}_8^{ib_1\cdots a_4b_4}
    \underline{\epsilon}_8^{c_1d_1\cdots c_4d_4} +
  {h_i}^{b_1}\underline{\epsilon}_8^{a_1i\cdots a_4b_4}
    \underline{\epsilon}_8^{c_1d_1\cdots c_4d_4} ).
\end{align}
It is notable how similar this is to the expansion of $t_8t_8$. With a
slight abuse of notation, the above can be represented as
\begin{equation} \label{eq:t8e8similarity}
  \epsilon_8^{abcdefgh} \to \underline{\epsilon}_8^{abcdefgh}
  -2( {h_i}^a\underline{\epsilon}_8^{ibcdefgh}
  +{h_i}^b\underline{\epsilon}_8^{aicdefgh} ),
\end{equation}
which is a direct analogue of \eqref{eq:t8exp}.

Due to this similarity between the $t_8t_8$ and
$\epsilon_{10}\epsilon_{10}$ expansions, the expansions of
$\epsilon_{10}\epsilon_{10}R^4$ and
$\epsilon_{10}\epsilon_{10}D^{2n}R^4$ are essentially identical to
those for $t_8t_8$ in sections \ref{sec:R4exp} and
\ref{sec:D6R4exp}. However, due to the extra antisymmetry of
$\epsilon_{10}$, there are further simplifications. In particular, the
four-vertex from $\epsilon_{10}\epsilon_{10}R^4$ or
$\epsilon_{10}\epsilon_{10}D^{2n}R^4$ expanded to lowest order
vanishes since the $\epsilon_{10}$ version of \eqref{eq:R4exp} is a
total derivative. Of course, the expansion to the next order does give
rise to a non-trivial five-vertex, which is why the presence of
$\epsilon_{10}\epsilon_{10}R^4$ and
$\epsilon_{10}\epsilon_{10}D^{2n}R^4$ can be studied using the
five-graviton amplitude.

Despite the fact that the four-vertex vanishes, the similarity between
the $t_8t_8$ and $\epsilon_{10}\epsilon_{10}$ expansions makes it
economical to ignore this fact so that the $t_8t_8$ results of the
previous section can be reused in almost unchanged form. In
particular, the diagrams of section \ref{sec:R4diagrams} are
practically identical to the equivalent
$\epsilon_{10}\epsilon_{10}R^4$ diagrams, simply with $t_8t_8$
replaced by $\frac{1}{4}\epsilon_8\epsilon_8$. As a consequence, the
matching of the amplitude with the effective action proceeds exactly
as in sections \ref{sec:R4Matching}-\ref{sec:D6R4Matching}.

At order $R^4$ the analysis mirrors that in section
\ref{sec:R4Matching}: the $\epsilon_8\epsilon_8$ parts of the
amplitude match with the $\epsilon_{10}\epsilon_{10}R^4$ effective
action diagrams which, of course, they must since there are no
conceivable terms to correct for a discrepancy. This confirms the
presence of the $\epsilon_{10}\epsilon_{10}R^4$ term which previously
had been detected from a covariant RNS calculation. As mentioned
above, there is an unresolved issue since \eqref{eq:eeamp} appears to
give the wrong sign for the one-loop $\epsilon_{10}\epsilon_{10}$
parts of \eqref{eq:IIBeps} and \eqref{eq:IIAeps}. At the next order
the amplitude vanishes, implying the absence of both
$\epsilon_{10}\epsilon_{10}D^2R^4$ and $\epsilon_{10}\epsilon_{10}R^5$
terms. Recall that $\epsilon_{10}\epsilon_{10}R^5$ is shorthand for a
pair of contractions between the $\epsilon_{10}$ tensors. Terms with
only one contraction cannot be seen in the light-cone gauge, and terms
with no contractions do not contribute to the five-graviton
amplitude. At order $D^4R^4$ the amplitude again vanishes, showing
that there are no $\epsilon_{10}\epsilon_{10}D^4R^4$ and no
$\epsilon_{10}\epsilon_{10}D^2R^5$ terms.

Finally, at order $D^6R^4$, the amplitude is non-zero and, as for the
$t_8t_8$ terms, requires both $D^6R^4$ and $D^4R^5$ terms in the
effective action. The analysis proceeds as in section
\ref{sec:D6R4Matching}, with the important $\sum\bar{t}_{10}=0$
identity continuing to hold for $\epsilon_8$ tensors as explained in
appendix \ref{sec:tensors}. From \eqref{eq:D6R4term}, the equivalent
$\epsilon_{10}\epsilon_{10}D^6R^4$ term is
\begin{align}
  S_{D^6\Rf} &= \pm\zeta(3)\zeta(2) \alpha'^6 \int
  d^{10}x\sqrt{-g}\,
  e^{-\phi}\,{\epsilon_{10\,mn}^{}}^{a_1b_1a_2b_2a_3b_3a_4b_4}
  \epsilon_{10}^{mnc_1d_1c_2d_2c_3d_3c_4d_4} \notag\\
  &\qquad\times  D_eD_fD_gR_{a_1b_1c_1d_1}
  D^eD^fD^g R_{a_2b_2c_2d_2}
  R_{a_3b_3c_3d_3} R_{a_4b_4c_4d_4}.
\end{align}
If it is correct to resolve the IIA/IIB sign issue at order $R^4$
simply by swapping the signs, then it is likely that the same is also
true at this order. So we postulate that the $+/-$ sign applies to
IIB/IIA respectively. Similarly, the
$\epsilon_{10}\epsilon_{10}D^4R^5$ terms are given by
\eqref{eq:D4R5terms} but with $t_8t_8$ again replaced by
$\pm\frac{1}{8}\epsilon_{10}\epsilon_{10}$.

So, in conclusion, in the same way that the usual one-loop $t_8t_8R^4$
term is replaced by
$(t_8t_8\pm\frac{1}{8}\epsilon_{10}\epsilon_{10})R^4$, the higher order
terms studied here -- $D^2R^4$, $R^5$, $D^4R^4$, $D^2R^5$, $D^6R^4$,
$D^4R^5$ -- require the same modification. This leads to
the obvious conjecture that one-loop $t_8t_8$ and
$\epsilon_{10}\epsilon_{10}$ tensors always appear in the combination
$(t_8t_8\pm\frac{1}{8}\epsilon_{10}\epsilon_{10})$ at all orders. Of
course, there are likely to be other $\epsilon_{10}\epsilon_{10}$ terms
with fewer contractions between the epsilon tensors, but these cannot
be analysed with this formalism. It is also natural to speculate that
the tree-level $D^{2n}R^4$, $R^5$ and $D^{2n}R^5$ terms may appear
multiplied by the combination
$(t_8t_8+\frac{1}{8}\epsilon_{10}\epsilon_{10})$ for both IIA and IIB,
as is the case for $R^4$.

%%%%%%%%%%%%%%%%%%%%%%%%%%%%%%%%%%%%%%%%%%%%%%%%%%%%%%%%%%%%%%%%%%%%%%%%%%%%%%%%%%%%%%%%%%%%%%%%%%%%%%%%%%%%%%
\section{Conclusions} \label{sec:conc}
%%%%%%%%%%%%%%%%%%%%%%%%%%%%%%%%%%%%%%%%%%%%%%%%%%%%%%%%%%%%%%%%%%%%%%%%%%%%%%%%%%%%%%%%%%%%%%%%%%%%%%%%%%%%%%
We considered the one-loop five-graviton amplitude in type II string
theory expanded up to order $D^6R^4$ and inferred various consequences
for the effective action. In particular, we determined the exact form
of the one-loop $D^{2n}R^4$ terms and whether new $R^5$ and
$D^{2n}R^5$ corrections are required.

Using the light-cone gauge it is not possible to determine all
terms. In particular, $\epsilon_{10}\epsilon_{10}$ terms with fewer than
two contractions, such as \eqref{eq:ee1contraction}, will always be
missed. However, all other terms should be visible. Since
$D^2R_{abcd}$ can be exchanged for a sum of $R^5$ terms as in
\eqref{eq:D2RisRR}, there is often ambiguity in the precise meaning of
$D^{2n}R^m$. To avoid confusion we write all terms using the maximum
possible number of Riemann tensors, so that $D^{2n}R^m$ refers to all
$D^{2n}R^m$ terms which cannot be rewritten using more Riemann
tensors.

At order $\alpha'^4$ relative to the Einstein-Hilbert term, it was
found that no $D^2R^4$ (which can be rewritten as a particular sum of
$R^5$ terms) or other $R^5$ terms are required. At the next order, we
confirmed the vanishing of $D^4R^4$ at one-loop, which was already
known from the four-graviton amplitude, and showed the absence of
$D^2R^5$ terms.

Up to this order no $R^5$ or $D^{2n}R^5$ terms are required. However,
at order $\alpha'^6$ the $D^6R^4$ term is not sufficient to account
for the five-graviton amplitude: new $D^4R^5$ terms are needed. It was
found that the most economical definition for $D^6R^4$ is as in
\eqref{eq:D6R4term}, with the additional $D^4R^5$ terms given by
\eqref{eq:D4R5terms}.

The famous $t_8t_8R^4$ term is extended in more complete treatments to
$(t_8t_8\pm\frac{1}{8}\epsilon_{10}\epsilon_{10})R^4$, with $+$ for
IIB and $-$ for IIA. Modulo the issue below, we were able to confirm
this structure at order $R^4$ and show that it extends to higher
orders: all the terms studied here -- $D^2R^4$, $R^5$, $D^4R^4$,
$D^2R^5$, $D^6R^4$, $D^4R^5$ -- were found to have the same behaviour,
with all non-zero terms being multiplied by a
$(t_8t_8\pm\frac{1}{8}\epsilon_{10}\epsilon_{10})$ factor. However,
there is an unresolved issue relating to the sign of the
$\epsilon_{10}\epsilon_{10}$ term. For both IIA and IIB, the
light-cone gauge GS formalism seems to predict the opposite sign to
that which is known to be true from covariant RNS calculations and
from $SL(2,\mathbb{Z})$ considerations in type IIB.

It is striking that, up to order $D^6R^4$, the coefficient in front of
the $D^{2n-2}R^5$ term matches with the coefficient in front of the
$D^{2n}R^4$ term. This leads to various five-graviton {\it
tree-level} conjectures which involve the same kinematic structure as
at one-loop but with coefficients from the tree-level four graviton
expansion. For IIB these conjectures can be extended to all orders
(even non-perturbatively): perhaps the modular function in front of
$D^{2n}R^4$ also appears in front of $D^{2n-2}R^5$. In fact this
modular function may be universal for all terms at the same order, not
just $D^{2n}R^4$ and $D^{2n-2}R^5$, but also $D^{2n-4}R^6$,
$D^{2n-6}R^7$, and so on. It would be useful to study amplitudes
involving more gravitons in order to test this.

\acknowledgments
I am very grateful to Michael Green, Hugh Osborn and Pierre Vanhove
for many useful discussions.

\appendix

%%%%%%%%%%%%%%%%%%%%%%%%%%%%%%%%%%%%%%%%%%%%%%%%%%%%%%%%%%%%%%%%%%%%%%%%%%%%%%%%%%%%%%%%%%%%%%%%%%%%%%%%%%%%%%
\section{Relationships Between $t_{10}$, $\bar{t}_{10}$ and $t_8$ Tensors} \label{sec:tensors}
%%%%%%%%%%%%%%%%%%%%%%%%%%%%%%%%%%%%%%%%%%%%%%%%%%%%%%%%%%%%%%%%%%%%%%%%%%%%%%%%%%%%%%%%%%%%%%%%%%%%%%%%%%%%%%
In addition to the usual eight-index $t_8$ tensor encountered in the
four-graviton amplitude, two new ten-index tensors, $t_{10}$ and
$\bar{t}_{10}$, appear in the five-graviton amplitude. These can both
be written as sums of $\delta t_8$ tensors. $\bar{t}_{10}$ appears
when the $S$ trace involves eight zero modes and two non-zero modes,
and can be rewritten using $t_8$ as
\begin{equation} \label{eq:bart10tot8}
  \bar{t}_{10}^{\,abcdefghij} = - \delta^{ad}t_8^{bcefghij} -
  \delta^{ac}t_8^{dbefghij} - \delta^{bc}t_8^{adefghij} -
  \delta^{bd}t_8^{caefghij}.
\end{equation}
From the symmetry of $t_8$, it is easy to show that
$\bar{t}_{10}^{\,a_1b_1a_2b_2\cdots}$ is antisymmetric under switching
$a_r$ with $b_r$, antisymmetric under switching $(a_1,b_1)$ with
$(a_2,b_2)$, and symmetric under switching $(a_s,b_s)$ with
$(a_t,b_t)$ for $\{s,t\}\in\{3,4,5\}$.

The $t_{10}$ tensor arises as the trace over five $R_0^{ab}$ tensors,
\eqref{eq:t10}, and can be expressed using $t_8$ as
\begin{align} \label{eq:t10tot8}
   2t_{10}^{a_1a_2\cdots a_{10}} &=
    -\;\delta^{a_1a_4}t_8^{a_2a_3a_5a_6a_7a_8a_9a_{10}}
    -\delta^{a_2a_3}t_8^{a_1a_4a_5a_6a_7a_8a_9a_{10}}
    +\delta^{a_1a_3}t_8^{a_2a_4a_5a_6a_7a_8a_9a_{10}} \notag\\
   &\quad\,+\delta^{a_2a_4}t_8^{a_1a_3a_5a_6a_7a_8a_9a_{10}}
    -\delta^{a_1a_6}t_8^{a_2a_5a_3a_4a_7a_8a_9a_{10}}
    -\delta^{a_2a_5}t_8^{a_1a_6a_3a_4a_7a_8a_9a_{10}} \notag\\
   &\quad\,+\delta^{a_1a_5}t_8^{a_2a_6a_3a_4a_7a_8a_9a_{10}}
    +\delta^{a_2a_6}t_8^{a_1a_5a_3a_4a_7a_8a_9a_{10}}
    -\delta^{a_1a_8}t_8^{a_2a_7a_3a_4a_5a_6a_9a_{10}} \notag\\
   &\quad\,-\delta^{a_2a_7}t_8^{a_1a_8a_3a_4a_5a_6a_9a_{10}}
    +\delta^{a_1a_7}t_8^{a_2a_8a_3a_4a_5a_6a_9a_{10}}
    +\delta^{a_2a_8}t_8^{a_1a_7a_3a_4a_5a_6a_9a_{10}} \notag\\
   &\quad\,-\delta^{a_1a_{10}}t_8^{a_2a_9a_3a_4a_5a_6a_7a_8}
    -\delta^{a_2a_9}t_8^{a_1a_{10}a_3a_4a_5a_6a_7a_8}
    +\delta^{a_1a_9}t_8^{a_2a_{10}a_3a_4a_5a_6a_7a_8} \notag\\
   &\quad\,+\delta^{a_2a_{10}}t_8^{a_1a_9a_3a_4a_5a_6a_7a_8}
    -\delta^{a_3a_6}t_8^{a_4a_5a_1a_2a_7a_8a_9a_{10}}
    -\delta^{a_4a_5}t_8^{a_3a_6a_1a_2a_7a_8a_9a_{10}} \notag\\
   &\quad\,+\delta^{a_3a_5}t_8^{a_4a_6a_1a_2a_7a_8a_9a_{10}}
    +\delta^{a_4a_6}t_8^{a_3a_5a_1a_2a_7a_8a_9a_{10}}
    -\delta^{a_3a_8}t_8^{a_4a_7a_1a_2a_5a_6a_9a_{10}} \notag\\
   &\quad\,-\delta^{a_4a_7}t_8^{a_3a_8a_1a_2a_5a_6a_9a_{10}}
    +\delta^{a_3a_7}t_8^{a_4a_8a_1a_2a_5a_6a_9a_{10}}
    +\delta^{a_4a_8}t_8^{a_3a_7a_1a_2a_5a_6a_9a_{10}} \notag\\
   &\quad\,-\delta^{a_3a_{10}}t_8^{a_4a_9a_1a_2a_5a_6a_7a_8}
    -\delta^{a_4a_9}t_8^{a_3a_{10}a_1a_2a_5a_6a_7a_8}
    +\delta^{a_3a_9}t_8^{a_4a_{10}a_1a_2a_5a_6a_7a_8} \notag\\
   &\quad\,+\delta^{a_4a_{10}}t_8^{a_3a_9a_1a_2a_5a_6a_7a_8}
    -\delta^{a_5a_8}t_8^{a_6a_7a_1a_2a_3a_4a_9a_{10}}
    -\delta^{a_6a_7}t_8^{a_5a_8a_1a_2a_3a_4a_9a_{10}} \notag\\
   &\quad\,+\delta^{a_5a_7}t_8^{a_6a_8a_1a_2a_3a_4a_9a_{10}}
    +\delta^{a_6a_8}t_8^{a_5a_7a_1a_2a_3a_4a_9a_{10}}
    -\delta^{a_5a_{10}}t_8^{a_6a_9a_1a_2a_3a_4a_7a_8} \notag\\
   &\quad\,-\delta^{a_6a_9}t_8^{a_5a_{10}a_1a_2a_3a_4a_7a_8}
    +\delta^{a_5a_9}t_8^{a_6a_{10}a_1a_2a_3a_4a_7a_8}
    +\delta^{a_6a_{10}}t_8^{a_5a_9a_1a_2a_3a_4a_7a_8} \notag\\
   &\quad\,-\delta^{a_7a_{10}}t_8^{a_8a_9a_1a_2a_3a_4a_5a_6}
    -\delta^{a_8a_9}t_8^{a_7a_{10}a_1a_2a_3a_4a_5a_6}
    +\delta^{a_7a_9}t_8^{a_8a_{10}a_1a_2a_3a_4a_5a_6} \notag\\
   &\quad\,+\delta^{a_8a_{10}}t_8^{a_7a_9a_1a_2a_3a_4a_5a_6},
\end{align}
from which it is clear that $t_{10}^{a_1b_1\cdots a_5b_5}$ is
antisymmetric under $a_r\leftrightarrow b_r$.

These tensors satisfy two important identities, both first discovered
in \cite{Frampton:1986ea}. The first relates the two ten-index tensors
by
\begin{align} \label{eq:t10identity}
2t_{10}^{abcdefghij} &= \bar{t}_{10}^{\,abcdefghij} +
\bar{t}_{10}^{\,abefcdghij} + \bar{t}_{10}^{\,abghcdefij} +
\bar{t}_{10}^{\,abijcdefgh} + \bar{t}_{10}^{\,cdefabghij} \notag\\
&\quad + \bar{t}_{10}^{\,cdghabefij} +
\bar{t}_{10}^{\,cdijabefgh} + \bar{t}_{10}^{\,efghabcdij} +
\bar{t}_{10}^{\,efijabcdgh} + \bar{t}_{10}^{\,ghijabcdef},
\end{align}
which is easily shown using \eqref{eq:t10tot8} and
\eqref{eq:bart10tot8}.

The second involves only the $\bar{t}_{10}$ tensor,
\begin{equation} \label{eq:bart10identity}
  \bar{t}_{10}^{\,abcdefghij} + \bar{t}_{10}^{\,abefcdghij}
  + \bar{t}_{10}^{\,abghcdefij} + \bar{t}_{10}^{\,abijcdefgh} = 0,
\end{equation}
which is equivalent to the vanishing of a particular sum of sixteen
$\delta t_8$ tensors. The proof proceeds in two parts: the proof for
the sixty $\delta\delta\delta\delta$ terms in $t_8$ and the proof for
the $\epsilon_8$ term. In the first case the identity becomes a
sum of $16\times 60$ $\delta\delta\delta\delta\delta$ tensors which
cancel out in pairs. The proof for the $\epsilon$ terms is less
obvious, but follows from the eight-dimensional version of the
two-dimensional identity $\delta^{ab}\epsilon^{cd} +
\delta^{ac}\epsilon^{db} + \delta^{ad}\epsilon^{bc} = 0$. This is
easily seen to be true since it is manifestly antisymmetric in $b$,
$c$ and $d$. Equivalently, it expresses the fact the there are no
non-vanishing three-forms in two dimensions. Finally, it can be
understood as showing that three vectors in two dimensions are
necessarily linearly dependent. The eight-dimensional version
\cite{Frampton:1986ea} reads
\begin{align} \label{eq:8DGram}
 \delta^{ab}\epsilon_8^{cdefghij} + \delta^{ac}\epsilon_8^{defghijb} +
 \delta^{ad}\epsilon_8^{efghijbc} + \delta^{ae}\epsilon_8^{fghijbcd} +
 \delta^{af}\epsilon_8^{ghijbcde} & \notag\\
 + \delta^{ag}\epsilon_8^{hijbcdef} +
 \delta^{ah}\epsilon_8^{ijbcdefg} + \delta^{ai}\epsilon_8^{jbcdefgh} +
 \delta^{aj}\epsilon_8^{bcdefghi} &= 0,
\end{align}
and, if applied twice, readily verifies \eqref{eq:bart10identity} for
the $\epsilon_8$ tensors.

\bibliographystyle{utcaps}
\bibliography{references}

\end{document}